\documentclass[aps,prb,twocolumn,groupedaddress,showpacs]{revtex4}

\newlength{\myw}        
\newlength{\mycwidth}       
\usepackage{bm,dcolumn,graphicx,subfigure,latexsym,amssymb}
\begin{document}

\title{Broad boron sheets and boron nanotubes:\\An \textit{ab initio} study of structural, electronic, and mechanical properties}
\author{Jens Kunstmann}
\email{j.kunstmann@fkf.mpg.de}
\affiliation{Max-Planck-Institut f\"ur Festk\"orperforschung, Heisenbergstra{\ss}e 1, 70569 Stuttgart, Germany}

\author{Alexander Quandt}
\affiliation{Institut f\"ur Physik der Universit\"at Greifswald, Domstra{\ss}e 10a, 17489 Greifswald, Germany}

\date{\today}

\begin{abstract}
Based on a numerical \textit{ab initio} study, we discuss a structure model for a broad boron 
sheet, which is the analog of a single graphite sheet, and the precursor of boron nanotubes.  
The sheet has linear chains of $sp$ hybridized $\sigma$ bonds lying only along its armchair 
direction, a high stiffness, and anisotropic bonds properties.
The puckering of the sheet is explained as a mechanism to stabilize the $sp$ $\sigma$ bonds.
The anisotropic bond properties of the boron sheet lead to a two-dimensional reference 
lattice structure, which is rectangular rather than triangular. As a consequence the 
chiral angles of related boron nanotubes range from $0^{\circ}$ to $90^{\circ}$.
Given the electronic properties of the boron sheets, we demonstrate that all of the related 
boron nanotubes are metallic, irrespective of their radius and chiral angle, and we also 
postulate the existence of helical currents in ideal chiral nanotubes.
Furthermore, we show that the strain energy of boron nanotubes will depend on their radii, as 
well as on their chiral angles. This is a rather unique property among nanotubular systems, 
and it could be the basis of a different type of structure control within nanotechnology.
\end{abstract}

%
%
%
%
\pacs{81.07.De, 73.63.Fg, 61.46.+w}

\keywords{boron sheets, boron nanotubes, nanotechnology, ab initio simulations}

\maketitle


\section{\label{intro}Introduction}
Boron is an electron deficient element \cite{pauling_1960_ncb}
which has a rather fascinating chemistry. Pure boron compounds
neither have a purely covalent nor a purely metallic character.
This results in a chemical versatility, which is unique among the
elements of the periodic table.

The classical bulk modifications of boron are based on B$_{12}$
icosahedra. The simplest boron phase is rhombohedral
$\alpha$-boron, \cite{fujimori_1999_prl} where boron icosahedra
are centered on the edges of a rhombohedral unit cell. A different
picture arises for boron clusters, where quasiplanar isomers turn
out to be more stable than their icosahedral counterparts. This is
the consequence of an Aufbau principle for elemental boron
clusters postulated by Boustani. \cite{boustani_1997_prb} This
Aufbau principle generally states that stable boron clusters can
be constructed from two basic units only: a pentagonal pyramidal
B$_6$ unit and a hexagonal pyramidal B$_7$ unit,
\cite{boustani_1997_prb} and it implies quasiplanar,
\cite{boust2:97} tubular, \cite{boustani:97,gindulyte_1998_ic1}
convex, and spherical \cite{boustani_1997_jssc} boron clusters. The
existence of quasiplanar clusters or "sheets" was recently
confirmed by experiment, \cite{zhai_2003_natmat} in perfect
agreement with earlier theoretical predictions.
\cite{boustani_1999_cpl} Furthermore, the existence of quasiplanar
boron clusters implies the formation of boron nano\-tubes and/or
boron fullerenes, because during synthesis, a growing
(quasi-)planar boron cluster tends to remove dangling bonds by
forming closed tubular or polyhedral modifications. And indeed,
recent experimental studies demonstrate the existence of boron
nanotubes. \cite{ciuparu_2004_jpcb,kiran_2005_pnas}

Carbon nanotubes \cite{iijima:91} on the other hand are a structural 
paradigm for all nanotubular materials and they can be seen as 
cylindrical modi\-fications of graphite, which may geometrically be
constructed by cutting a rectangular piece out of a single
graphene sheet and rolling it up to form a tube. Almost all
properties of carbon nanotubes can be derived from the properties
of a single graphene sheet, which means that a profound
understanding of graphite is the key to understand the basic
properties of carbon nanotubes. The same relation holds for boron
sheets (BSs) and boron nanotubes (BNTs): understanding the
structure and the properties of BSs will be crucial for our
understanding of the basic properties of BNTs.

This paper builds on previous work \cite{boustani:99,quandt_2005_cpc} 
to establish such a basic connection between BSs and BNTs, but it 
should be pointed out that our previous reasoning was mainly based 
on the individual structures of finite sized quasiplanar boron clusters.
\cite{boust2:97,boustani:98,boustani:00} Using \textit{ab initio}
structural optimization methods for solid systems we could finally
discriminate among different structure models for layered boron
compounds and establish a simple model for a broad and stable BS.

After a detailed description of this search process, we will
analyze the properties of the most stable structure model. Then we
will show how these results may be used to explain the structure,
the stability, the electronic and the mechanical properties of
BNTs. In particular the somewhat surprising constriction of zigzag
BNTs, which has been reported in a recent publication,
\cite{kunstmann_2005_cpl} may now be clearly understood on the
basis of the elastic properties of BSs.

It must be pointed out that up to now, a \textit{broad} BS, which
would be the analog of a single graphite sheet, could not be found
experimentally. But when writing up this paper we became aware of
an interesting work by Evans \textit{et al.},
\cite{evans_2005_prb} who consider three BS models and five BNTs
of small tube radii, and the work of Cabria \textit{et al.} \cite{cabria_2006_nt} 
who study two BS models and three BNTs. Although our results are certainly based 
on a more extensive search for stable BS and BNTs, our findings for the stable BS 
are, from a structural and energetic point of view, in excellent agreement with 
these authors. Thus the present structure model could independently be confirmed 
by three different groups. However, there is still some disagreement about the 
ground state structures of BNTs. Lau \textit{et al.}, \cite{lau_2006_cpl} for example, 
have recently reported about structures for BS and BNTs, which are very different 
from the structure models of Evans \textit{et al.} and Cabria \textit{et al.}, but 
the present study is in clear favor of the latter.


\section{\label{sec:theory}Methods}

As pointed out by Pauling \cite{pauling_1960_ncb} elemental boron has a 
complicated and rather versatile chemistry. Therefore the only reliable 
theoretical tools, which may allow for a proper description of boron 
chemistry, are first principles calculations. \cite{boustani_1997_prb}

In order to carry out structural optimizations of BSs and BNTs we
used the \textsc{VASP} package, version 4.4.6.
\cite{kresse:96-1,kresse:96-2} The latter is a density functional
theory \cite{kohn:65} based {\em ab initio} code using plane wave
basis sets and a supercell approach to model solid materials,
surfaces, or clusters. \cite{payne_1992_rmp} During all of our
simulations, the electronic correlations were treated within the
local-density approximation (LDA) using the
Perdew-Zunger-Ceperley-Alder exchange-correlation functional,
\cite{perdew_1981_prb,ceperley:80} and the ionic cores of the
system were represented by ultrasoft pseudopotentials
\cite{vanderbilt:90} as supplied by Kresse and Hafner.
\cite{kresse_1994_jpcm} The $k$-space integrations were carried
out using the method of Methfessel and Paxton
\cite{methfessel_1989_prb} in first order, where we employed a
smearing width of 0.3 eV.

With the help of the \textsc{VASP} program, one can determine interatomic
forces, which may be used to relax the different degrees of
freedom for a given decorated unit cell. Eventually one will
detect some atomic configurations, which correspond to (local) minima
on the total energy landscape. In order to carry out those
extensive structure optimizations in a more effective way, we employed a
conjugate gradient algorithm, \cite{payne_1992_rmp} and we allowed all of 
the atomic coordinates to relax, as well as all but one lattice
parameter. This rigid lattice parameter would fix the interlayer
separation for BS and the intertubular distance for BNTs at 6.4
\AA, which effectively makes them stand-alone objects. The sizes of 
the $k$-point meshes for different systems with different unit cells 
were individually converged, such that changes in the total energy were 
reduced to less than 3 meV/atom. In the course of a structural 
optimization run, all interatomic forces were finally reduced to less 
than 0.04 eV/\AA. The cutoff energy for the expansion of the electronic 
wave functions in terms of plane waves was 257.1 eV for the relaxation 
runs, and 321.4 eV for a final static calculation of the total energy.

The cohesive energies given in Tables \ref{tab:lattice} and
\ref{tab:BNT} were calculated from
\begin{equation}
E_{\mathrm{coh}} = E_{\mathrm{bind}}/n \label{eqn:Ecoh}.
\end{equation}
$E_{\mathrm{bind}}$ is the the atomic binding energy per unit cell
and $n$ is the number of atoms per unit cell. Therefore in our
definition $E_{\mathrm{coh}}$ will be a positive number.

For band structures and the analysis of Fermi surfaces in Sec.~\ref{sec:prop_elec} 
and Appendix \ref{app:flat} we used the Stuttgart TB-LMTO-ASA package, which 
is a density functional theory \cite{kohn:65} based code using short range
\cite{andersen_1984_prl} linearized muffin-tin orbitals
\cite{andersen_1975_prb} within the atomic sphere approximation
(ASA). It allows static calculations of the electronic properties
for periodic systems. We used the non-spin polarized LDA
exchange-correlation functional of Barth and Hedin
\cite{barth_1972_jpc} and a $k$-mesh of 30 x 30 x 3.


\begin{figure}[t]
\centering
\includegraphics[width=0.9\mycwidth]{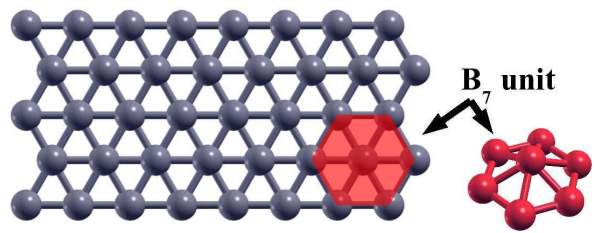}
\caption{\label{fig:sheet}(Color online) Top view of a quasiplanar boron 
sheet.
In a planar projection the atoms form an almost perfect triangular
lattice. The basic structural unit is a hexagonal pyramidal B$_7$
cluster, as suggested by the Aufbau principle
(Ref.~\onlinecite{boustani_1997_prb}) (see text).}
\end{figure}

\section{\label{sec:BS}Boron Sheets}

\subsection{\label{sec:model}Finding a structure model}
Following the Aufbau principle \cite{boustani_1997_prb} a BS is
basically a quasiplanar arrangement of hexagonal pyramidal B$_7$
units. A planar projection of such a system will always form some
kind of triangular lattice (see Fig.~\ref{fig:sheet}). However, the
out of plane modulation (i.e., the puckering) remains unspecified
by the Aufbau principle. The latter has to be determined using
\textit{ab initio} structural optimizations, after setting up a
suitable supercell that will allow for a systematic generation of
various periodic puckering schemes.

The versatile chemistry of boron is reflected in a complicated
energy landscape, which is full of local minima. Therefore the
standard optimization techniques like the conjugate gradients
method used in this study are most likely to find local minima,
rather than global minima. Therefore we examined the energy
landscape quite carefully by performing \textit{many} optimization
runs, which started from quite diverse initial configurations.

\begin{figure}[t]
\setlength{\myw}{3cm} 
\centering
\subfigure(a){\includegraphics[height=\myw]{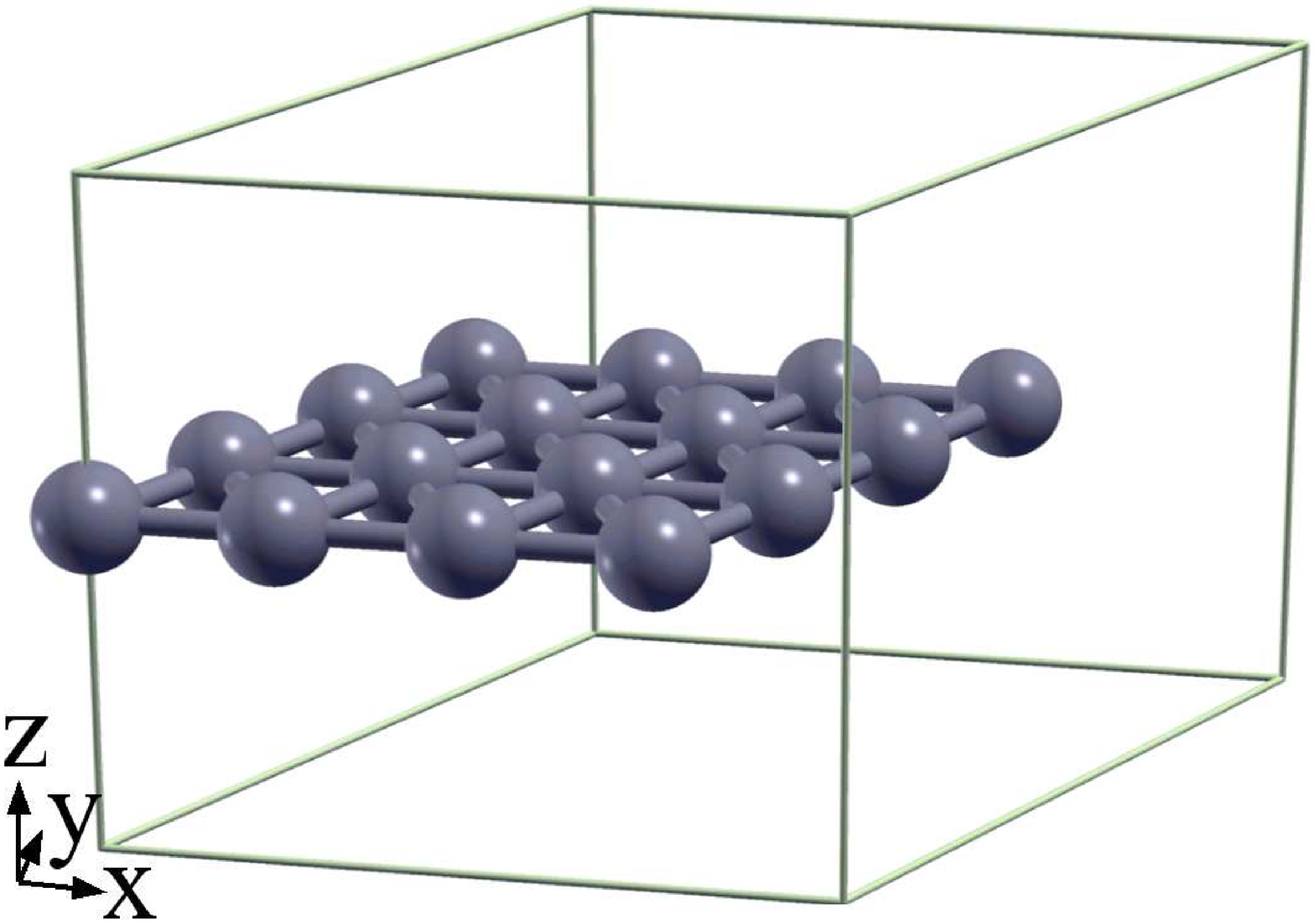}}
\subfigure(b){\includegraphics[height=\myw]{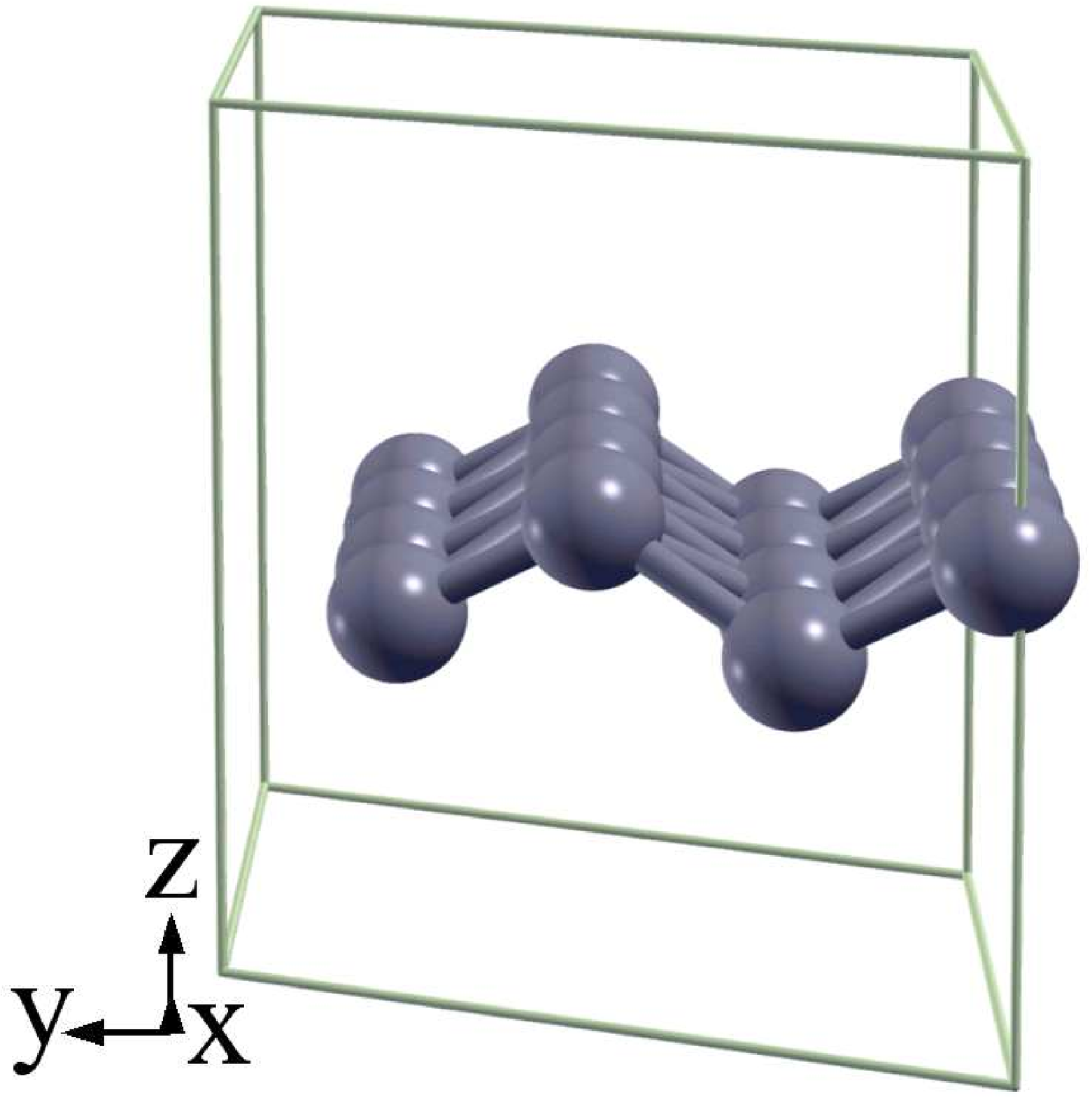}}
\subfigure(c){\includegraphics[height=\myw]{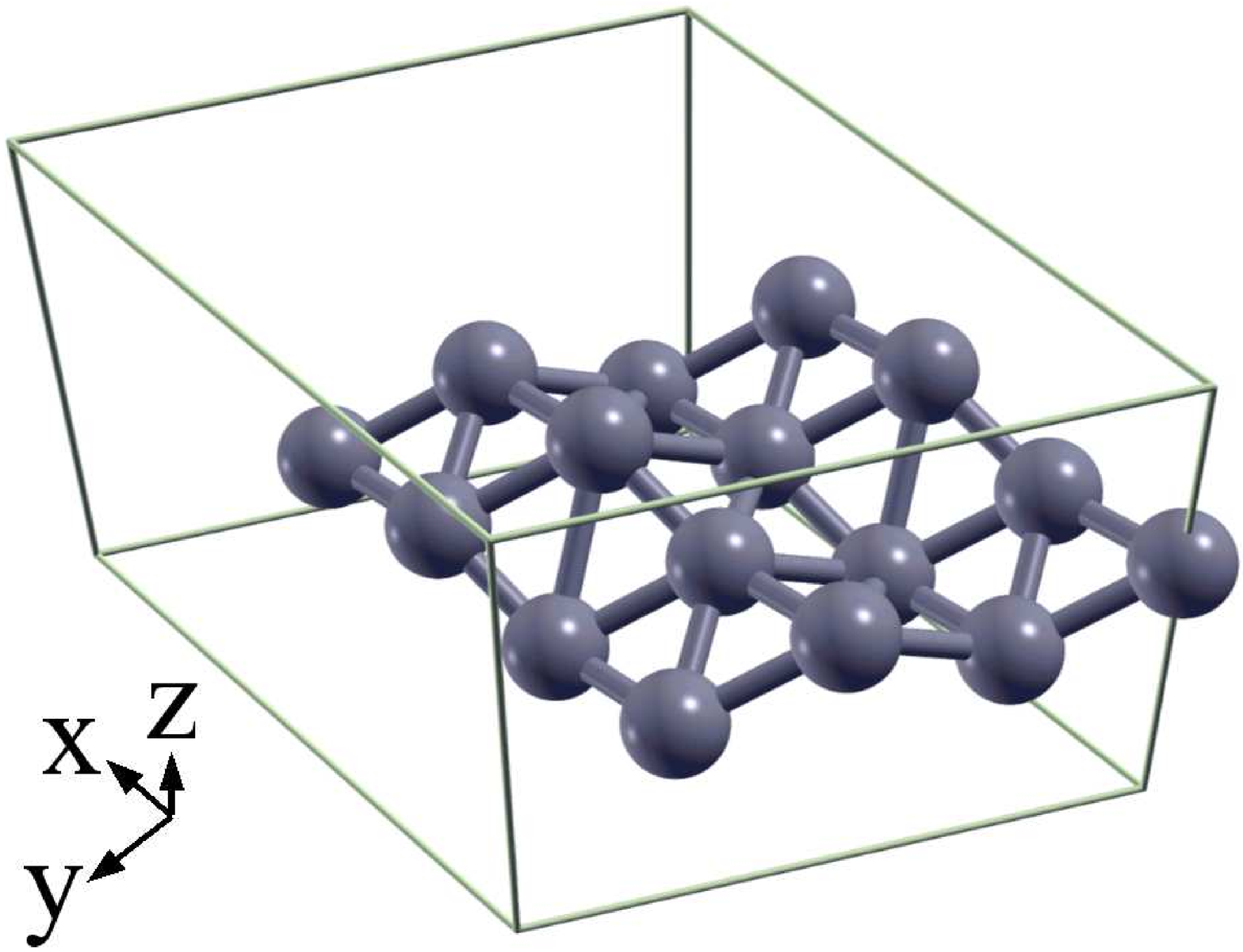}}
\subfigure(d){\includegraphics[height=\myw]{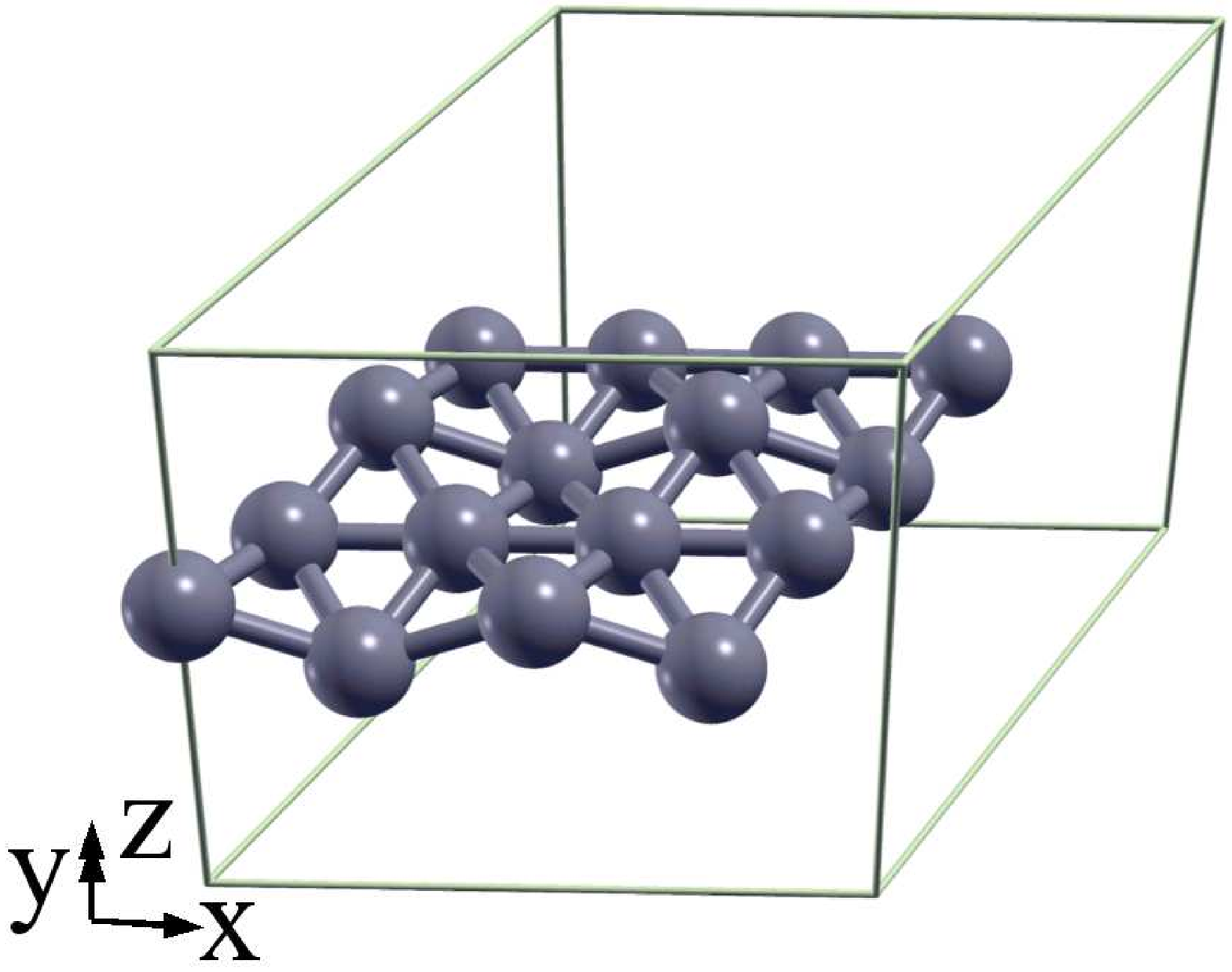}}
\caption{\label{fig:puck}(Color online) Different structure models
for broad boron sheets. Each supercell (thin lines) contains 16
atoms. (a) A simple flat sheet is metastable. (b) A simple up and
down puckering seems to be the most stable modulation. Structures
(c) and (d) are unstable. Models (b), (c), and (d) are periodic
repetitions of structural motives taken from B$_{22}$, B$_{46}$,
and B$_{32}$ clusters, described in Ref.~\onlinecite{boust2:97}. }
\end{figure}

The basic puckering schemes were taken from the structures of
B$_{22}$, B$_{32}$, and B$_{46}$ clusters, which are described in
Ref.~\onlinecite{boust2:97}. We repeated the puckering
periodically in a triangular supercell containing 16 atoms (see
Figs.~\ref{fig:puck}(b)-\ref{fig:puck}(d)), and optimized the resulting
structures. For the sake of comparison we also examined a flat BS
(see Fig.~\ref{fig:puck}(a)). \footnote{The initial in-plane
boron-boron distance was 1.6 \AA. The $k$-point mesh was 4x4x2 for
the optimization runs and 6x6x4 for a final static calculation.}
The flat boron sheet (a) occupies a local minimum on the energy
landscape with a cohesive energy of 6.76 eV/atom, but small
out-of-plane elongations of individual atoms immediately cause a
puckering of the BS. This was confirmed by shifting one atom 0.1,
0.2, and 0.4 {\AA} out of plane and reoptimizing the resulting
structures. Thus model (a) turns out to be metastable (as also pointed
out in Refs.~\onlinecite{evans_2005_prb} and \onlinecite{cabria_2006_nt}); 
any thermal vibration would lead to a permanent deformation of a flat boron 
sheet. Models (c) and (d) are completely unstable, and they both relax to
structure (b). In order to scan the energy landscape for other
candidate structures we took sheet (a) and shifted each of the 16
atoms out of the plane, employing a random elongation $\Delta z$
between $+$0.4 and $-$0.4 {\AA}. Those structures were reoptimized
as before. It turns out that 8 out of 11 optimizations led
to model (b), while the remaining three runs resulted in a
metastable kinked structure with a cohesive energy of 6.86 eV/atom
(see Appendix \ref{app:kink}).

The fact that models (c) and (d) as well as 8 out of 11
randomly puckered sheets would relax to model (b) means that
structure (b) defines a rather pronounced minimum on the energy
landscape. The high structural stability of model (b) is confirmed
by its high cohesive energy of 6.94 eV/atom, which is the highest
cohesive energy of all BSs that we found. We thus conclude that
the most suitable structure model for a broad BS will be (b),
being 0.18 eV/atom more stable (0.21 and 0.26 eV/atom in 
Refs.~\onlinecite{cabria_2006_nt} and \onlinecite{evans_2005_prb}, 
respectively) than an unrealistic flat BS.
The puckering itself seems to be an important
mechanism to stabilize the BS, \cite{cabria_2006_nt} which will be examined in more
detail in Sec.~\ref{sec:props}.

\begin{table}[b]
\caption{\label{tab:lattice} Detailed LDA description of the
optimized lattice structures of the flat (a) and puckered (b)
boron sheets (see Figs.~\ref{fig:puck} and \ref{fig:cell}), their
bond lengths, cohesive energies $E_{\mathrm{coh}}$ (Eq.
(\ref{eqn:Ecoh})), and their elastic moduli $C_x=C_{11}$ and
$C_y=C_{22}$ obtained after stretching a sheet along the Cartesian
$x$ or $y$ direction (Eqs.~(\ref{eqn:modx}) and (\ref{eqn:mody})).}
 \begin{ruledtabular}
 \begin{tabular}{l|ll}
 Sheet & (a) Flat & (b) Puckered\\
 \hline
Lattice type& Triangular (2D) & Orthorhombic (3D)\\
Lattice param. (\AA)& $A=1.69$ & $A=2.82$\\
&& $B=1.60$\\
&& $C=$ arbitrary\\
Primitive vectors & $\bm{a}_1= A(\frac{\sqrt{3}}{2},\frac{1}{2})$ 
& $\bm{a}_1=A(1,0,0)$\\
& $\bm{a}_2=A(\frac{\sqrt{3}}{2},-\frac{1}{2})$ & $\bm{a}_2=B(0,1,0)$\\
&& $\bm{a}_3=C(0,0,1)$\\
Atoms/unit cell & 1 & 2 \\
Atomic pos. (\AA)& $\bm{R}_1 = (0,0)$ & $\bm{R}_1 = (0,0,0)$\\
&& $\bm{R}_2 = (\frac{1}{2}A,\frac{1}{2}B,0.82)$\\
 \hline
Bond lengths (\AA) & $a_{\mathrm{B-B}}=1.69$ & 
$a_{\mathrm{B-B}}^{\sigma}=1.60$ \\
&& $a_{\mathrm{B-B}}^{\mathrm{diagonal}}=1.82$ \\
\hline
$E_{\mathrm{coh}}$ (eV) & $6.76$ & $6.94$\\
\hline
Elastic modulus & $C_x = C_y = 0.75$ & $C_x = 0.42$\\
(TPa)&&$C_y = 0.87$\\
 \end{tabular}
 \end{ruledtabular}
\end{table}

\begin{figure}[t]
\centering
\includegraphics[width=0.5\mycwidth]{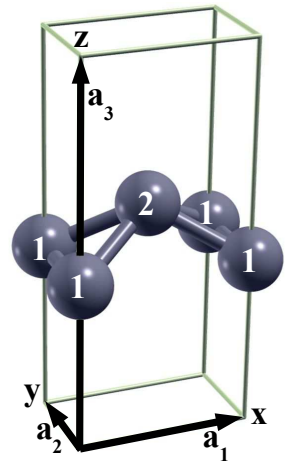}
\caption{\label{fig:cell}(Color online) The orthorhombic unit cell of model 
(b) with two basis atoms (see Table \ref{tab:lattice}). In a
$xy$-projection atom 1 is located at the corners of a rectangular
unit cell, while atom 2 is located at the center of the unit cell.
Along the $z$ direction the boron atoms will generate a simple up
and down puckering, with puckering heights around $\Delta z =
0.82$ {\AA}.}
\end{figure}

In order determine the lattice structures of (a) and (b) we
performed LDA calculations, where we would fix the
unit cell of each system for a series of Cartesian lattice
constants $A$ or $B$, whereas all of the internal (atomic) degrees
of freedom were allowed to relax. The resulting total energies for
a given set of lattice constants were fitted to polynomial curves
$E(A)$ and $E(B)$, from which we determined the equilibrium
properties of the systems. The results are summarized in
Table \ref{tab:lattice}.
The diagonal elements of the elastic tensor $C_x=C_{11}$ and
$C_y=C_{22}$ may be interpreted as a first approximation to a
macroscopic Young's modulus. They were calculated as follows:

\begin{eqnarray}
C_x = \frac{A_0}{Bh} \left( \frac{\partial^2E(A)}{\partial A^2} \right)_{A_0}, 
\label{eqn:modx}\\
C_y = \frac{B_0}{Ah} \left( \frac{\partial^2E(B)}{\partial B^2} \right)_{B_0}, 
\label{eqn:mody}
\end{eqnarray}
$h$ is the height of the BS, and it was defined as $h = \Delta z +
2 R_{\mathrm{vdW}}$; $\Delta z$ is the puckering height of the
sheet and $R_{\mathrm{vdW}}$ is the van der Waals radius.
\footnote{ The definition of $h$ looks somewhat arbitrary, as a
different definition for $h$ or different van der Waals radii will
certainly alter the values for the elastic moduli. But in a test
calculation for a single graphite sheet, where we used $\Delta z =
0$ (no puckering) and $R_{\mathrm{vdW}}^{\mathrm{C}} = 1.7$ {\AA},
we found $C_{11}=C_{22}$ to be 1.08 TPa, in excellent agreement
with the literature values of $C_{11}=1.06$ TPa.
\cite{saito_1989_cnt} For boron we would used
$R_{\mathrm{vdW}}^{\mathrm{B}} = 1.7$ {\AA} and $\Delta z = 0.82$
(see below and Table \ref{tab:lattice}). Thus for model (a) we find
that $h=3.4$ {\AA}, whereas for model (b) we find that $h=4.22$
{\AA}.} $A_0$ and $B_0$ are the equilibrium lattice constants.

The optimized planar model (a) seems to form a triangular lattice with 
one atom per unit cell and a single lattice constant $A$, which is in 
the range of a typical boron-boron bond length $A=a_{\mathrm{B-B}}=1.69$ 
\AA. But within the accuracy of the given methods, we cannot really 
decide whether the lattice structure is perfectly triangular or slightly 
less symmetric. Assuming perfect triangular symmetry the two elastic
moduli $C_x$ and $C_y$ are equal, and they are
surprisingly big: $C_y=C_y \approx 750$ GPa. Which means that even
if the flat BS is metastable compared to other model boron sheets,
it nevertheless has an extraordinary high stiffness. In Appendix
\ref{app:flat} we will analyze model (a) in more detail.

In Fig.~\ref{fig:cell} we depicted the unit cell of model (b). It
consists of two basis atoms, and its planar projection is almost
triangular, but not quite so.
It is common to describe such a system with a face centered
rectangular unit cell with lattice constants $A$ and $B$. For $A/B =
\sqrt{3} = 1.732$ a planar projection of the system would be
equivalent to a triangular system. In our case $A/B = 1.76$, which is
a small, but noticeable departure from triangular symmetry. Due to
a puckering height of $\Delta z = 0.82$ \AA, such a system might
best be described using a three-dimensional orthorhombic unit
cell. The corresponding lattice parameters and bond lengths can be
found in Table \ref{tab:lattice}.
\footnote{$a_{\mathrm{B-B}}^{\mathrm{diagonal}}=
a_{\mathrm{B-B}}^{1-2}$ is the bond length between atom 1 and 2,
and $a_{\mathrm{B-B}}^{\sigma} = a_{\mathrm{B-B}}^{1-1} =
a_{\mathrm{B-B}}^{2-2} = B$ is the bond length between two
equivalent atoms in different unit cells.}


\subsection{\label{sec:props}Properties of the model boron sheet}

In this section we will analyze the properties of model (b), which
turns out to be the most stable structure for broad BS. Therefore,
whenever we write 'boron sheet' (BS) in the following, we will
only refer to model (b).

In order to compare the BS with a known boron structure we also
calculated the cohesive energy of the $\alpha$ boron, which turns out
to be 7.51 eV/atom. This corresponds to an energetic difference of
0.57 eV/atom (0.58 and 0.57 eV/atom in Refs.~\onlinecite{evans_2005_prb} and 
\onlinecite{cabria_2006_nt}, respectively), which is huge, but one has to 
take into account that we are comparing a single boron sheet with a bulk 
reference structure.

\subsubsection{Mechanical properties}

\begin{figure}[t]
\centering
\includegraphics[width=0.95\mycwidth]{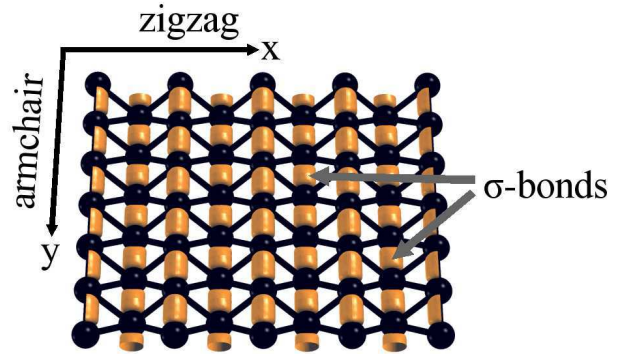}
\caption{\label{fig:sigma} (Color online) Orange (gray): charge density 
contours of the boron sheet (model (b)) at 0.9 e/\AA$^3$. One observes
parallel linear chains of $sp$ hybridized $\sigma$ bonds lying
along the armchair direction. }
\end{figure}

The elastic modulus of model (b) strongly depends on the
stretching directions. In Table \ref{tab:lattice} we roughly find
that $C_y \approx 2 C_x$. How can one explain those rather obvious
anisotropies?

To this end, let us have a look at the charge density of the BS
(see Fig.~\ref{fig:sigma}). We clearly observe some parallel
linear chains of $\sigma$ bonds lying along the armchair
direction. Their bond length is $a_{\mathrm{B-B}}^{\sigma} = 1.60$
\AA. At lower densities ($\rho < 0.7$ e/\AA$^3$, not displayed) a
largely homogeneous distribution with a rather complex shape
appears, which may be assigned to multicenter bonding typical for
boron materials. An analysis of the electron localization function
\cite{becke_1990_jcp} (ELF) leads to similar results, such that we
obtain the following preliminary picture of the bonding: on a
first level the sheet is held together by homogeneous multicenter
bonds, but on a second level there are strong $\sigma$ bonds lying
only along the armchair direction.

Due to the strong $\sigma$ bonds, any stretching of the BS along
the armchair ($=y$) direction will be much harder than a similar
stretching along its zigzag ($=x$) direction, where only the
slightly weaker multicenter bonds are involved.
These results are quite different from the results of Evans
\textit{et al.}, who conjecture that the $\sigma$ bonds are strong
but soft. \cite{evans_2005_prb} But here we clearly observe that the
$\sigma$ bonds are strong and stiff.
However other basic findings of Evans \textit{et al.}
are in good agreement with our results for flat and puckered BSs.

In general the elastic moduli involved are quite high; the stiffness 
of the $\sigma$ bonds along the armchair direction is comparable to 
the stiffness of a graphene sheet. Furthermore the broken triangular 
symmetry of the BS's 2D lattice structure, as mentioned in Sec.
\ref{sec:model}, is another immediate consequence of the
anisotropic bond properties.

Evans \textit{et al.} also found that BNTs of different
chiralities have different stiffnesses. \cite{evans_2005_prb} This
can be confirmed by our bonding picture, although our results
suggest that zigzag BNTs should be somewhat stiffer than armchair
BNTs, while Evans \textit{et al.} noted the opposite (the
armchair and zigzag direction are swapped in their and our
treatment, see Appendix \ref{app:wvector}).
We thus conclude that the
relation between the microscopic elastic modulus and the
macroscopic Young's modulus must be rather complicated in the case
of BS and BNTs.

\subsubsection{\label{sec:prop_elec}Electronic properties}

\begin{figure}[t]
\centering
\includegraphics[angle=-90, width=0.95\mycwidth]{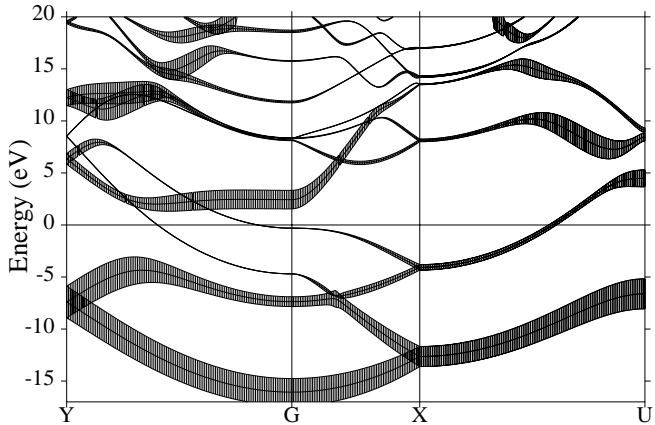}
\caption{\label{fig:bands} The band structure of the model BS.
The fatness of the bands indicates their $sp$ character, and it shows
that the $\sigma$ bonds in Fig.~\ref{fig:sigma} must be of
$sp$ type. The Fermi energy $E_F$ lies at $E=0$, $G$ is the $\Gamma$
point. }
\end{figure}

The two-dimensional band structure of the BS
$E^{\mathrm{BS}}(k_x,k_y)$ is plotted in Fig.~\ref{fig:bands}
along lines of high symmetry. The BS is metallic, as there are
two bands crossing the Fermi energy, which is in perfect agreement
with earlier studies of BSs. \cite{boustani:98,boustani:99}

In order to find out about the hybridization of the $\sigma$
bonds, we plotted the corresponding amount of $s$ and $p_y$
character indicated by the fatness of the bands. \footnote{The
orientation of the $p_x$, $p_y$, and $p_z$ orbitals coincides with
the orientation of the coordinate systems in Figs.~\ref{fig:cell}
and \ref{fig:sigma}.} We do not find individual dispersions of $s$
or $p$ bands, and the lowest lying bands show dispersions which
\textit{share} $s$ and $p_y$ character. That means they are bands
consisting of $sp$ hybridized orbitals:
\begin{eqnarray*}
|sp_a \rangle &=& \frac{1}{\sqrt{2}} ( |s \rangle + |p_y \rangle )\\
|sp_b \rangle &=& \frac{1}{\sqrt{2}} ( |s \rangle - |p_y \rangle ).
\end{eqnarray*}
The directional coincidence of the $p_y$ orbitals with the
$\sigma$ bonds in Fig.~\ref{fig:sigma} identifies them to be of
$sp$ type. The strength of the $\sigma$ bonds originates from the
fact that the bands lie 5 to 15 eV below the Fermi energy.

The physical picture to describe the multicenter bonds seems to
be much more complicated and it is still under investigation (see
Appendix \ref{app:flat}). So far we tried to analyze the
multicenter bonds using a simple tight binding model, which
comprises the remaining $p_x$ and $p_z$ orbitals as basis states.
But it turned out that this treatment can only partially reproduce
the conduction bands in Fig.~\ref{fig:bands}; probably a larger
basis set is needed.

In Sec. \ref{sec:model} we indicated that the puckering has a
stabilizing effect for the BS. Now we are in a good
position to explain this observation: any flattening of the BS
would cause $p_x$ orbitals to interfere with the $\sigma$ bonds
and eventually destroy them. An analysis of the charge density
and ELF of a flat BS (see Appendix \ref{app:flat}) indeed shows
that there are no $\sigma$ bonds involved, but only multicenter bonds.

The existence of $sp$ rather than $sp^2$ hybridization in a quasi
twodimensional layered structure is somewhat surprising. Earlier
studies of quasiplanar boron clusters \cite{boust2:97,boustani:00}
still presumed the presence of $sp^2$ hybridization. We think that
these assumptions should be reconsidered.

\begin{figure}[t]
\centering
\includegraphics[width=0.66\mycwidth]{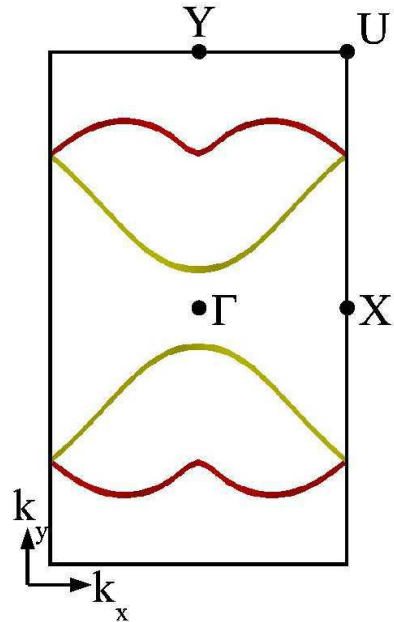}
\caption{\label{fig:FS}(Color online) The two-dimensional Fermi surface of 
the boron sheet. It consists of two contours in red (black) and yellow (gray), 
which correspond to the two bands crossing the Fermi energy in 
Fig.~\ref{fig:bands}. }
\end{figure}

Finally we want to discuss the two-dimensional Fermi surface
$E_F=E^{\mathrm{BS}}(k_x,k_y)$ of the BS in Fig.~\ref{fig:FS}. It
obviously consists of two contours, which are dispersed throughout
the Brillouin zone. This clearly shows the metallic properties of
the BS.

\begin{table*}
\caption{\label{tab:BNT} Structural data and energies of different
isomers of free standing boron nanotubes: $(k,l)$, $(n,m)$, $(i,j)$:
different chiral indices for the same tube type (see Appendix
\ref{app:wvector}); $n$: number of atoms per unit cell; Isom.:
label of isomer; $C_j$: rotational symmetry;
$a_{\mathrm{B-B}}^{\mathrm{axial}}$,
$a_{\mathrm{B-B}}^{\mathrm{diagonal}}$,
$a_{\mathrm{B-B}}^{\mathrm{circumferential}}$: boron-boron bond
lengths in axial, diagonal and circumferential direction of a
nanotube, the superscript $\sigma$ indicates that this bond is a
$\sigma$ bond, superscripts ${\sigma, i}$ and ${\sigma, o}$ refer to inner
and outer rings, respectively; $\bar{R} \pm \Delta R$: mean radius
of a nanotube (Eq.~(\ref{eqn:Rbar})) and maximal radial variation
(Eq.~(\ref{eqn:DeltaR})); $E_{\mathrm{coh}}^{\mathrm{ind}}$:
cohesive energy of a free standing (individual) nanotube 
(Eq. (\ref{eqn:Ecoh}));
$E_{\mathrm{coh}}^{\mathrm{rope}} -
E_{\mathrm{coh}}^{\mathrm{ind}}$: this energy is gained when the
same nanotube is arranged in a bundle (rope). All energies are given 
in eV/atom and all lengths are given in {\AA}.}

 \newcolumntype{e}[0]{D{/}{/}{5.7}}
 \begin{ruledtabular}
 \begin{tabular}{leccllllldd}
 $(k,l)$
 & (n,m)/(i,j)
 & $n$
 & Isom.
 & $C_j$
 & $a_{\mathrm{B-B}}^{\mathrm{axial}}$
 & $a_{\mathrm{B-B}}^{\mathrm{diagonal}}$
 & $a_{\mathrm{B-B}}^{\mathrm{circumferential}}$
 & $\bar{R} \pm \Delta R$
 & \multicolumn{1}{c}{$E_{\mathrm{coh}}^{\mathrm{ind}}$}
 & \multicolumn{1}{c}{$E_{\mathrm{coh}}^{\mathrm{rope}} -   
E_{\mathrm{coh}}^{\mathrm{ind}}$}\\
 \hline
(9,0) &(9,0)/(9,9) & 18& $\alpha$ & C$_3$ & 1.61$^{\sigma}$ & 1.77,1.83,1.86 && 3.86 $\pm$ 1.09 & 6.93 &+0.07\\ 
           &&& $\beta$ & C$_1$ & 1.61$^{\sigma}$ & 1.67 $-$ 1.87 &&& 6.92 & \\ 
                   &&& $\gamma$ & C$_3$ & 1.61$^{\sigma}$ & 1.81,1.82 && 3.83 $\pm$ 0.51 &6.91 & +0.04\\ 
           &&& $\delta$ & C$_9$ & 1.61$^{\sigma}$ & 1.83 && 4.17 $\pm$ 0.39 & 6.83 &\\ 
           &&& $\epsilon$ & C$_3$ & 1.64$^{\sigma}$ & 1.67,1.81 && 4.39 $\pm$ 0.29 & 6.78 &   \\ 
\\
(10,0)&(10,0)/(10,10)& 20 & $\alpha$& C$_2$ & 1.60$^{\sigma}$ & 1.79,1.81,1.82,1.87 && 3.84 $\pm$ 1.97  &6.91&+0.01\\ 
                   &&& $\beta$ & C$_2$ &1.61$^{\sigma}$ & 1.82,1.83,1.84 && 4.08 $\pm$ 1.18 &6.90 & +0.07\\ 
           &&& $\gamma$ & C$_{10}$ & 1.61$^{\sigma}$ & 1.83 && 4.60 $\pm$ 0.41 & 6.85 & \\ 
\\
(12,0)&(12,0)/(12,12)& 24 & $\alpha$ & C$_6$ & 1.61$^{\sigma}$ & 1.73,1.83,1.85 && 5.05 $\pm$ 0.65 &6.90&+0.02\\ 
                   &&& $\beta$ & C$_{12}$ & 1.61$^{\sigma}$ & 1.82 && 5.48 $\pm$ 0.41 & 6.87 & +0.05 \\ 

\\
\\

(0,12)&(4,4)/(12,0)& 24 &$\alpha$ & C$_6$ && 1.69 & 1.59$^{\sigma}$,1.69,1.85 & 2.64 $\pm$ 0.68 & 6.68 & +0.3\\ 
\\
(0,18)&(6,6)/(18,0)& 36 & $\alpha$ & C$_6$ && 1.70,1.74 & 1.56$^{\sigma}$,1.60$^{\sigma}$,1.71,1.75 & 4.48 $\pm$ 0.57 & 6.74 & +0.27\\ 
                  &&& $\beta$  & C$_{18}$ && 1.75 & 1.53$^{\sigma}$,1.76 &4.74 $\pm$ 0.34 &6.72 & \\  
\\
(0,24)&(8,8)/(24,0)& 48 & $\alpha$ & C$_6$ && 1.74,1.75 & 1.54$^{\sigma ,i}$,1.57$^{\sigma ,i}$,1.64$^{\sigma ,o}$,1.72,1.74 & 5.99 $\pm$ 0.58 &6.81 & +0.3\\ 
 \end{tabular}
 \end{ruledtabular}

\end{table*}


\section{\label{sec:BNT}Boron Nanotubes}

In Sec. \ref{sec:tubestruc} we will show that the structure
of BNTs is strongly related to the structure of the BS, such that
the latter may be seen as a direct precursor of BNTs. Therefore it
will be interesting to try to characterize BNTs simply by referring
to the BS. The mathematical details of such a relation are discussed
in Appendix \ref{app:idealBNT}, and when proceeding along these lines, 
a BNT may be characterized by two numbers $(k,l)$ with $k,l \ge 0$.

For the \textit{ab initio} simulation of BNTs we would start from a series
of initial structures with smooth surfaces, which were
optimized in a triangular BNT bundle (rope). Here the strong
tube-tube interactions (see Sec.~\ref{sec:ropes}) distort the surfaces 
and naturally induce some puckering. The energy of this configuration 
is $E_{\mathrm{coh}}^{\mathrm{rope}}$.
In order to simulate free standing (individual) BNTs we
would then increase the intertubular distance to 6.4 {\AA}, and
optimize those configurations again while keeping the
intertubular distances fixed. The energy here is 
$E_{\mathrm{coh}}^{\mathrm{ind}}$. ($E_{\mathrm{coh}}^{\mathrm{rope}}$ 
and $E_{\mathrm{coh}}^{\mathrm{ind}}$ are defined after Eq.~(\ref{eqn:Ecoh}))
\footnote{The $k$-space integration
for free standing BNTs was carried out on a 1x1x15 mesh for zigzag
tubes and on a 1x1x10 mesh for armchair types.}

All free standing BNTs are shown in Figs.~\ref{fig:zigzag-9},
\ref{fig:zigzag-10+12}, and  \ref{fig:armchair} and the structural
data and energies are collected in Table \ref{tab:BNT}. Besides
their bond lengths and rotational symmetries we also stated the
geometrical mean radius of each tube $\bar{R}$, as well as the
maximal radial variation $\Delta R$, which were defined as:
\begin{eqnarray}
\bar{R} &=& \frac{R^{\mathrm{min}} + R^{\mathrm{max}}}{2},
   \label{eqn:Rbar}\\
\Delta R &=& R^{\mathrm{max}} - \bar{R} = \bar{R} - R^{\mathrm{min}},
   \label{eqn:DeltaR}
\end{eqnarray}
where $R^{\mathrm{min}}$ and $R^{\mathrm{max}}$ are the distances
of the innermost and the outermost atoms from the center of the
nanotube, respectively.

For many $(k,l)$ BNTs we found more than just one isomer.
Therefore each BNT was also given a Greek index which labels different 
isomers. The latter were ordered according to their cohesive energies, 
i.e., $(k,l)\alpha$ will denote the most stable isomer, $(k,l)\beta$ would 
be less stable, and so on.


\subsection{\label{sec:ropes}Free standing nanotubes vs nanotube ropes}

In Table \ref{tab:BNT} the "inter-tubular energy"
$E_{\mathrm{coh}}^{\mathrm{rope}} -
E_{\mathrm{coh}}^{\mathrm{ind}}$ is the energetic difference
between a free standing BNT and its bundled counterpart. One can see that
it varies significantly from tube to tube. The intertubular energy seems to
depend quite strongly on the structure type, the relative orientations of
adjacent tubes in a rope, and the specific type of surface puckering.
Furthermore, the intertubular distance in different bundles, which was
defined as the minimal separation between two apex atoms on adjacent
nanotubes, varies between 1.7 and 3.5 {\AA} in our simulations.

It is obvious that the tube-tube interaction in BNT bundles
(ropes) is completely different from what is known from carbon
nanotubes, where the intertubular interaction is of van der Waals
type. The latter is certainly much weaker, independent of the various
structure types, and the intertubular distances are always around
3.4 {\AA}. BNTs on the other hand \textit{may} have
\textit{covalent} intertubular bonds, \cite{kunstmann_2005_cpl,
quandt:01} and this leads to a decent intertubular bonding energy
that depends quite strongly on structural details.

It is interesting to note that the intertubular energy of $(0,l)$
BNTs (armchair types) is significantly higher than for $(k,0)$
BNTs (zigzag). In Sec. \ref{sec:armchair} we will try to give
an explanation for this rather complex bonding scenario.

\begin{figure}[t]
\setlength{\myw}{5cm} 
\centering
\includegraphics[width=\mycwidth]{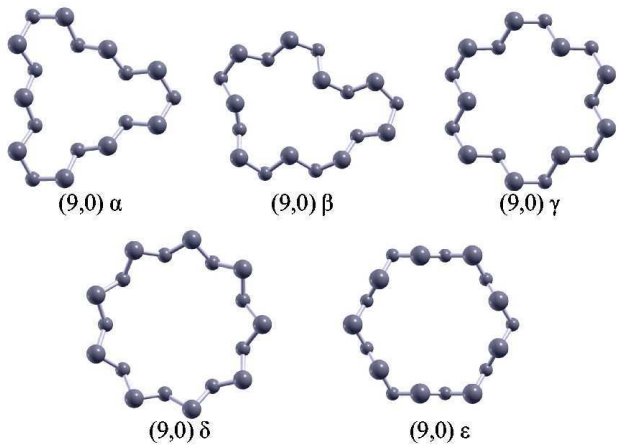}
\caption{\label{fig:zigzag-9}(Color online) The cross sections of different
isomers of a free standing (9,0) zigzag boron nanotube. The big
spheres stand for the upper atoms and the small ones for the lower
atoms (with respect to the direction of the tube axis). The
$\alpha$ and $\gamma$ isomers are the free standing counterparts
of the (9,0)C and (9,0)B tubes in
Ref.~\onlinecite{kunstmann_2005_cpl}, respectively.}
\end{figure}

\begin{figure}[t]
\centering
\includegraphics[width=0.9\mycwidth]{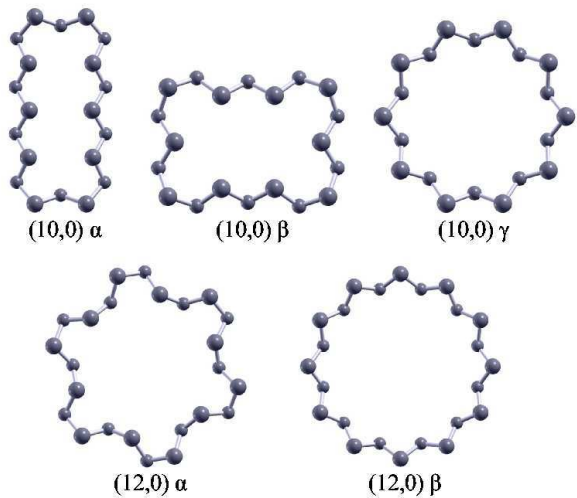}
\caption{\label{fig:zigzag-10+12}(Color online) Cross-sectional view of
various isomers of free standing (10,0) and (12,0) zigzag boron
nanotubes. Again the big spheres mark the upper atoms and the
small ones mark the lower atoms. The (10,0)$\alpha$ and
(10,0)$\beta$ isomers are free standing counterparts of the
(10,0)B and (10,0)C structures in
Ref.~\onlinecite{kunstmann_2005_cpl}, respectively.}
\end{figure}

At this point, it will be worth noting that the original motivation
for this paper was a recent study by ourselves, where we reported
bundled zigzag BNTs that were somewhat \textit{constricted}
\cite{kunstmann_2005_cpl} (we define the concept of \textit{constriction} 
at the end of Sec.~\ref{sec:zigzag}). We conjectured that this constriction
would most likely be caused by the arrangement of the tubes in a bundle,
where the tube-tube interactions will force the tubes to have
geometrical shapes different from free standing BNTs. Now the
free standing counterparts of the constricted (9,0)C and (10,0)C BNTs
from Ref.~\onlinecite{kunstmann_2005_cpl} are the
(9,0)$\alpha$ \footnote{For the (9,0)$\alpha$ BNT the
intertubular distance was increased to only 4 \AA.} isomer in
Fig.~\ref{fig:zigzag-9} and (10,0)$\beta$ in
Fig.~\ref{fig:zigzag-10+12}. To our surprise the constriction
would \textit{not} disappear after isolating the tube. And even
after substantially deforming the (9,0)$\alpha$ structure by
homogeneous shrinking, by blowing it up, or by randomly elongating
atoms out of their equilibrium position with a maximum amplitude of 0.3
{\AA}, the free standing (9,0)$\alpha$ BNTs would always relax to
their constricted forms. This finding is in clear contrast to our
previous hypothesis, and it raises the important question where those
constrictions finally come from. We will try to give an answer to
this question in Sec. \ref{sec:strain}.


\subsection{\label{sec:tubestruc}The structure of free standing boron 
nanotubes}

\begin{figure}[t]
\setlength{\myw}{3.8cm} 
\centering
\subfigure(a){\includegraphics[width=\myw]{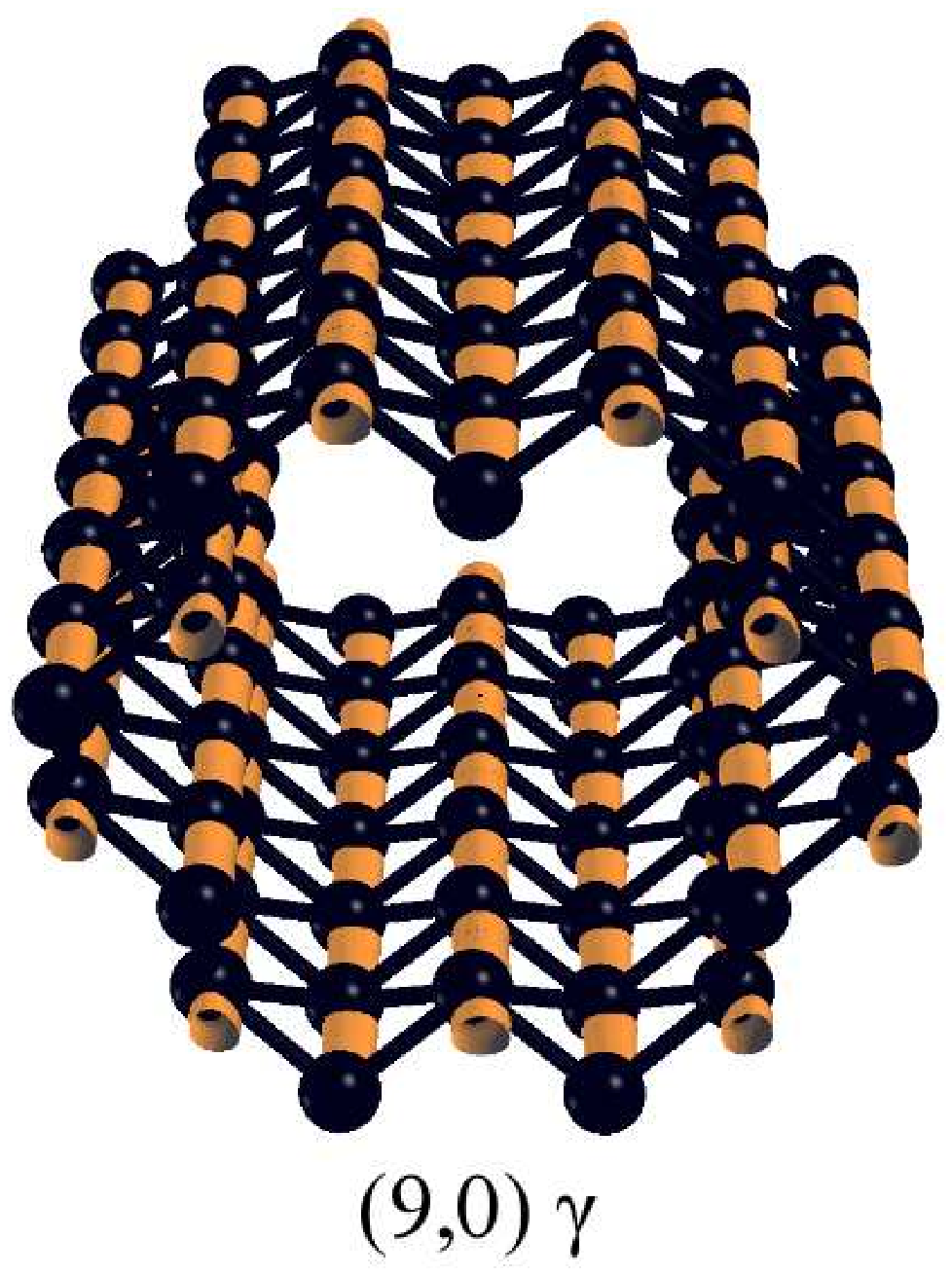}}
\subfigure(b){\includegraphics[width=\myw]{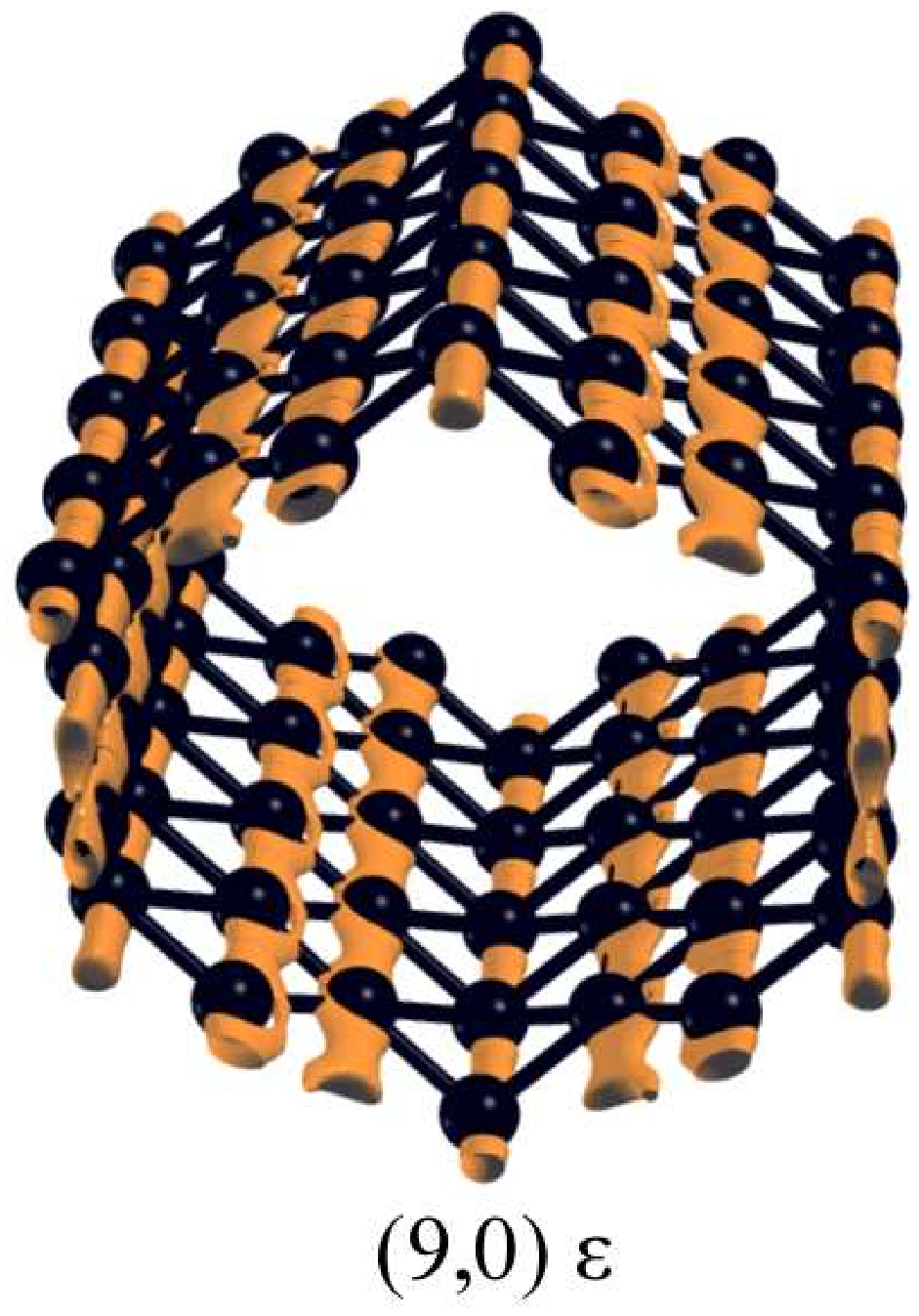}}
\caption{\label{fig:ChgZigzag}(Color online) Zigzag boron
nanotubes and the presence of straight $\sigma$ bonds along their
axial direction, which are indicated by orange (gray) charge
density contours at 0.9 e/\AA$^3$. Due to a lack of stiff
$\sigma$ bonds along the circumferential direction, this type of
nanotube might not be stable.}
\end{figure}

\begin{figure}[t]
\setlength{\myw}{5cm} 
\centering
\includegraphics[width=0.8\mycwidth]{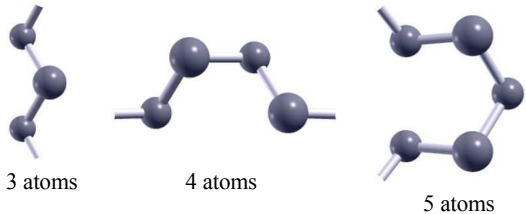}
\caption{\label{fig:elements}(Color online) Three basic structure elements. 
The cross sections of the most stable zigzag boron nanotubes in
Figs.~\ref{fig:zigzag-9} and \ref{fig:zigzag-10+12} may be composed
using these elements only. }
\end{figure}


\subsubsection{\label{sec:zigzag}Zigzag nanotubes}

\begin{figure*}[t]
\setlength{\myw}{8.5cm} 
\centering
\includegraphics[width=\linewidth]{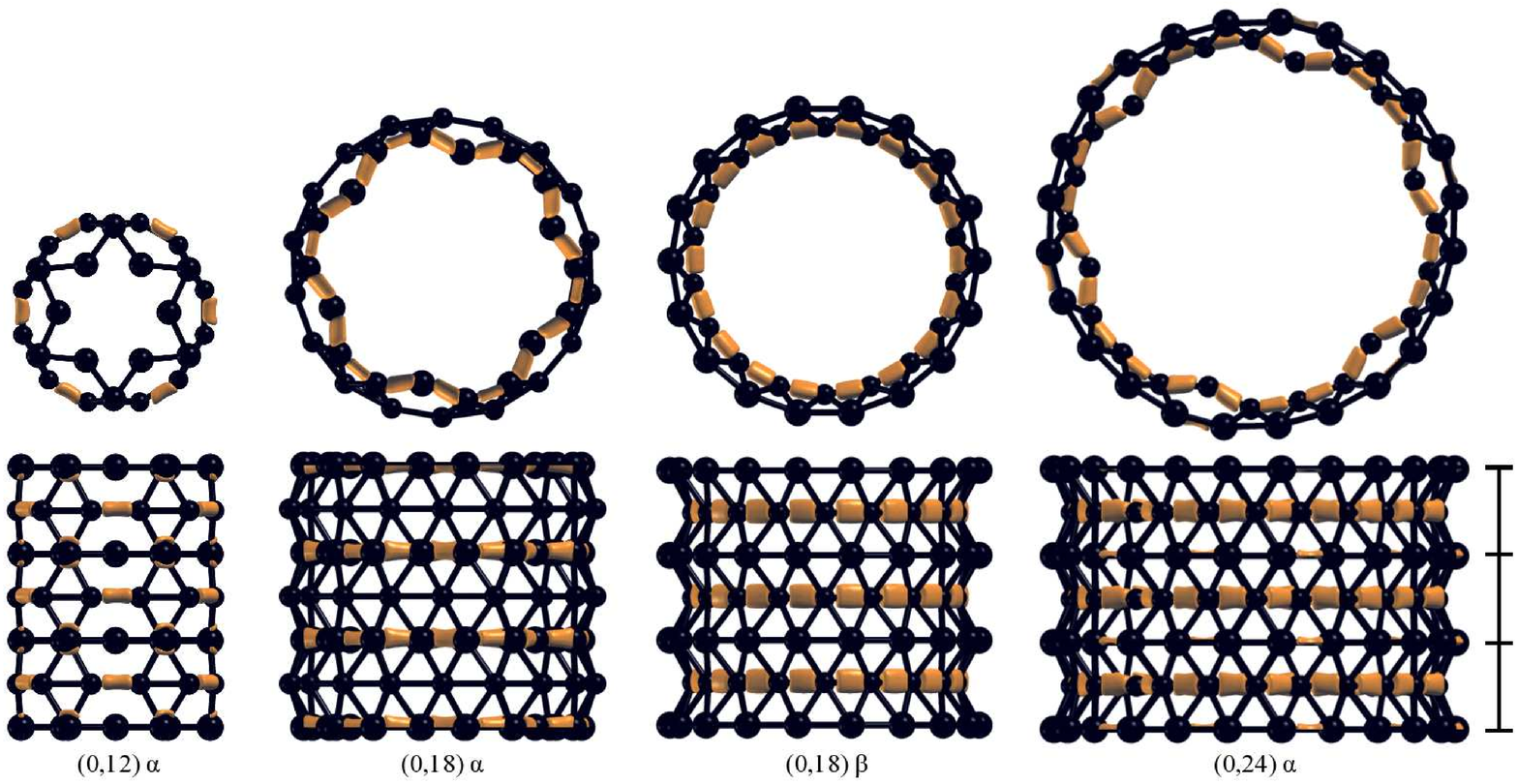}
\caption{\label{fig:armchair}(Color online) Top and side view of various free
standing armchair boron nanotubes and the presence of $\sigma$
bonds, which are indicated by orange (gray) charge density
contours at 0.95 e/\AA$^3$. All armchair nanotubes have bent
$\sigma$ bonds along the circumferential direction, which basically
generate the strain energies of the tubes. The black bar on the right
indicates the height of a supercell in axial direction that was used
for our simulations; for aesthetical reasons we actually displayed
three identical units cells.}
\end{figure*}

For zigzag BNTs we found various isomers. Any
zigzag BNT may be seen as a BS that was rolled up along its zigzag
direction (see Fig.~\ref{fig:sigma} or \ref{fig:cut&roll}). Thus the
linear chains of $\sigma$ bonds will lie along its axial direction
and they will remain straight. These basic bonding properties were
typical for \textit{all} zigzag BNT that we studied so far.
We just show two typical examples in Fig.~\ref{fig:ChgZigzag}. Here the 
bond length of the $\sigma$ bonds is quite similar to the bond length in 
the BS, as $a_{\mathrm{B-B}}^{\sigma} = a_{\mathrm{B-B}}^{\mathrm{axial}} =
1.61$ {\AA} (the only exception we found was (9,0)$\epsilon$).

The tubes (9,0)$\delta$, (10,0)$\gamma$, and (12,0)$\beta$ are \textit{ideal} 
BNTs, which denotes the fact that they were initially constructed by a cut and 
paste procedure described in Appendix \ref{app:wvector}, and then reoptimized 
using \textit{ab initio} methods. Their structure is highly symmetric and we find two 
bond lengths, which are almost identical to the bond length in the BS. The 
puckering height $\Delta z=2 \Delta R \approx 0.8$ is also quite similar to 
the BS.

However, an ideal BNT does not seem to be the ground state of a real zigzag BNT, 
and we found less symmetric isomers that were higher in cohesive energy. It 
should be noted that zigzag tubes with a smooth surface were not considered here 
because their cohesive energies are significantly lower than those of puckered 
BNTs. As an example for the complex shape of zigzag BNTs one may study 
(9,0)$\epsilon$, which is the least stable isomer of all zigzag BNTs. (9,0)$\epsilon$ 
has a hexagonal cross section, which probably arises from the triangular 
supercell into which it was put. Its sides may be seen as parts of a flat BS, 
whereas the corner pieces are parts of a puckered BS. From 
Fig.~\ref{fig:ChgZigzag} we notice that the $\sigma$ bonds along the sides are
slightly more delocalized than the ones located at the corners. This means that 
any flattening would destabilize the sigma bonds, and the whole tube is highly 
metastable. (A similar but squarelike structure was found by Evans 
\textit{et al.}, \cite{evans_2005_prb} which they labeled $(i,j)=(6,6)$, 
but we think that this structure is highly metastable as well.) Thus the 
question will no longer be if zigzag BNTs are puckered, but \textit{how} 
they are puckered.

The cross sections of the isomers with a high cohesive energy may be built from 
three basic structure elements that are shown in Fig.~\ref{fig:elements}. The 
three-atomic structure element is directly related to the puckering of the BS 
(compare Fig.~\ref{fig:sigma}) whereas the four- and five-atomic elements are 
just special combinations of three-atomic structure elements. We see that the 
structure of zigzag BNTs is strongly related to the \textit{local} structure 
of a simply puckered BS, but their \textit{general} cross-sectional geometries 
seem to be more complicated and less symmetric than in the simple case of an 
ideal BNT. This loss of symmetry can also be extracted from the spectrum of 
diagonal bond lengths, which are associated with multicenter bonds. Those 
bond lengths are not equal to $a_{\mathrm{B-B}}^{\mathrm{diagonal}}$ of the 
BS (see Table \ref{tab:lattice}), but span a whole range 
$a_{\mathrm{B-B}}^{\mathrm{diagonal}} \approx 1.7 - 1.9$ \AA.

Some of the most interesting structures are (9,0)$\alpha$, (9,0)$\beta$, 
and (10,0)$\alpha$, which have cross sections that are far from being 
circular. Nonetheless they exhibit high cohesive energies. Because of the 
observed unusual shapes of zigzag BNTs we assume that the multicenter bonds 
obviously possess a high directional flexibility, but at the same time they 
are also very stiff ($C_x=0.42$ TPa in Table \ref{tab:lattice}). Therefore it 
seems as if these bonds have some jointlike properties, i.e., they are easy to 
turn, but hard to tear.

In the following we will call a zigzag BNT \textit{constricted}, if it is 
composed of several five-atomic structure elements. In our work the 
(9,0)$\alpha$ and the (10,0)$\beta$ isomers are constricted. A constricted 
zigzag BNT was also found by Evans \textit{et al.} \cite{evans_2005_prb}
There it is labeled as a $(i,j)=(8,8)$ nanotube, and it corresponds to 
our (10,0)$\beta$ structure without the two horizontal three-atomic elements.

\subsubsection{\label{sec:armchair}Armchair nanotubes}

When rolling up a BS along its armchair direction, the puckered sheet 
(see Fig.~\ref{fig:sigma}) will be transformed into a tube that has inner and 
outer rings, and the $\sigma$ bonds will lie along its circumferential direction. 
On the outer rings the length of the $\sigma$ bonds will be increased and on the 
inner rings their length will be reduced. In Fig.~\ref{fig:armchair} we see
that for three systems discussed in this study, the $\sigma$ bonds do really 
lie along the circumferential direction, and for the (0,18) and (0,24) systems 
an inner and an outer ring can clearly be identified.

In contrast to zigzag BNTs, for the armchair types we did \textit{not} find 
several isomers, and we just discuss one ideal BNT, which is the (0,18)$\beta$ 
isomer. In analogy to zigzag BNTs, we found that this ideal BNT corresponds to 
a local energy minimum, and the (0,18)$\alpha$ isomer of lower symmetry is 0.02 
eV/atom more stable. \footnote{The $C_6$ symmetry of all $\alpha$ isomers is 
probably not an intrinsic property, but rather caused by the fact that they were 
simulated in a triangular supercell.} The latter has $\sigma$ bonds solely along 
the inner rings, where the bond lengths are 1.56 and 1.60 \AA. Along the outer 
ring, where the B-B distances (1.71 and 1.75 {\AA}) are significantly longer, 
the curvature effect has destroyed the $\sigma$ bonds. \footnote{The diagonal 
bond lengths $a_{\mathrm{B-B}}^{\mathrm{diagonal}}$ (which connect the inner
and the outer rings) are always shorter compared to the BS 
(compare Tables \ref{tab:lattice} and \ref{tab:BNT}); we found them to be 
in the range $a_{\mathrm{B-B}}^{\mathrm{diagonal}} \approx 1.69 - 1.75$ \AA.}

The (0,24) system has similar properties, but here the curvature is smaller, 
and there are six additional weak $\sigma$ bonds along the outer rings with a bond 
length of 1.64 \AA. For even larger radii we expect the outer rings of armchair 
BNTs to develop $\sigma$ bonds between every single atom.

The radius of the (0,12) BNT is quite small, which makes it extremely difficult 
for the structure to align its $\sigma$ bonds. We see that this tube possesses 
a different geometry, and even along the stiffer rings there are six instead of 
12 $\sigma$ bonds. It is obvious that for armchair BNTs with smaller and 
smaller radii, the curvature effect will successively destroy the circumferential 
$\sigma$ bonds. For the smallest possible BNTs there will probably be no 
$\sigma$ bonds at all, and the surface of the tube will become smooth. This 
agrees with earlier studies by ourselves \cite{boustani:97,boustani:98} and 
with the work of Evans \textit{et al.}, \cite{evans_2005_prb} where some 
armchair BNTs of small radii were studied and found to be smooth.

Any destruction of circumferential $\sigma$ bonds within armchair BNTs of small 
radii will release electrons that can alter their chemical properties. In 
Sec. \ref{sec:ropes} we observed that the intertubular energy for armchair 
BNT ropes is much higher than for zigzag BNT ropes. Now a possible explanation 
would be that the released electrons in armchair BNTs induce an enhanced
reactivity. In a rope of BNTs, this enhanced reactivity will lead to strong 
intertubular bonding for armchair BNTs of small radii. In zigzag BNTs the 
reactivity is lower, as a maximum number of $\sigma$ bonds can always be 
achieved, due to the fact that curvature effects will not be able to weaken 
the axial $\sigma$ bonds. Therefore we hypothesize that small sized armchair BNTs will have a higher 
reactivity than zigzag BNTs, and that this reactivity will further decrease 
with increasing radii.

This reactivity, which leads to the formation of intertubular bonds in BNT 
ropes, could be very useful when trying to embed BNTs into polymers, 
\cite{evans_2005_prb} where strong chemical bonds between the nanotubes 
and the polymer matrix are needed in order to improve the mechanical 
properties of the composite.


\subsection{\label{sec:strain}Strain energy}

\begin{figure}[t]
\centering
\includegraphics[width=0.9\mycwidth]{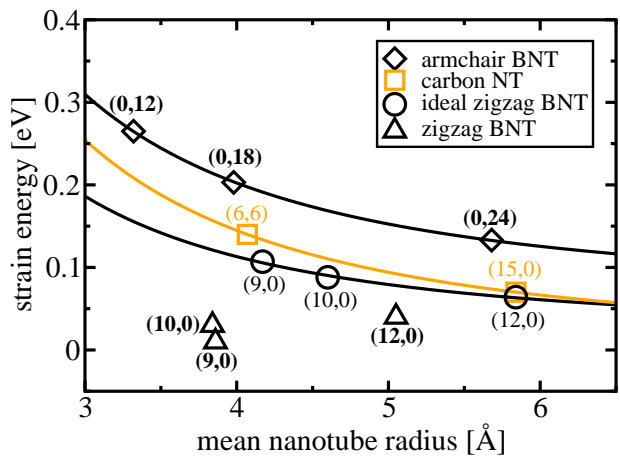}
\caption{\label{fig:strain}(Color online) The strain energy of $\alpha$ 
isomers as a function of the mean radius $\bar{R}$ (Eq.~(\ref{eqn:Rbar})); 
for armchair boron nanotubes we used $\bar{R}_{\sigma}$ 
(Eq.~(\ref{eqn:SigmaRbar})). In  orange (gray) we show the universal strain 
energy curve for carbon nanotubes ($\Box$); the energy obviously depends on 
their radii, but not on their chiral angles. For \textit{armchair} boron 
nanotubes ($\Diamond$) we find a similar curve, but those boron tubes have 
more strain energy. For \textit{zigzag} boron nanotubes ($\bigtriangleup$) 
we cannot really plot a strain energy curve, as different nanotubes of 
different radii are almost isoenergetic. \textit{Ideal} zigzag boron 
nanotubes ($\bigcirc$) have less strain energy than their armchair 
counterparts, but they are metastable.}
\end{figure}

Let us now compare the cohesive energy of every BNT 
($E_{\mathrm{coh}}^{\mathrm{ind}}$ from Table \ref{tab:BNT}) with the 
cohesive energy of the puckered BS ($E_{\mathrm{coh}}^{\mathrm{BS}}$
from Table \ref{tab:lattice}). This energy difference will be called 
strain energy:
\begin{equation}
E_{\mathrm{strain}}(k,l) = E_{\mathrm{coh}}^{\mathrm{BS}} - 
E_{\mathrm{coh}}^{\mathrm{ind}}(k,l).
\end{equation}
It is the amount of energy that is needed to roll up a BS into a BNT.
The microscopic origin of the strain energy in nanotubes are bent $\sigma$ 
bonds along the circumferential direction of the tubes. These bonds have a 
strong tendency to jump back into a straight orientation, which generates a 
tension that may be quantified by the strain energy of the systems. Such a 
tension will stabilize the tubular shape, or to put it more clearly: it will 
make the nanotube round. 

The strain energies of different $(k,l)$ BNTs as a function of their mean radii 
(Eq.~(\ref{eqn:Rbar})) is plotted in Fig.~\ref{fig:strain}. For the sake of 
comparison we also show the universal strain energy curve for carbon nanotubes. 
We call it universal because the strain energy only depends on the radius, 
but not on the chiral angle (chirality) of the nanotubes: 
$E_{\mathrm{strain}}^{\mathrm C}=E_{\mathrm{strain}}^{\mathrm C}(R)$.

As the BNTs are all puckered, there is some variability in the proper choice of 
a mean tubular radius. Since the strain energy is related to the position of the 
$\sigma$ bonds, it makes sense to define the mean radius of armchair BNTs as
\begin{equation}
\bar{R}_{\sigma} = \frac{R_{\sigma}^{\mathrm{min}} + 
R_{\sigma}^{\mathrm{max}}}{2}.
\label{eqn:SigmaRbar}
\end{equation}
Here $R_{\sigma}^{\mathrm{min}}$ and $R_{\sigma}^{\mathrm{max}}$
are the distances of the innermost and the outermost atoms sharing $\sigma$ 
bonds, which is measured from the center of the nanotube.

In earlier works we studied the elasticity of \textit{armchair} BNTs with a 
tight-binding method \cite{boustani:99} and reported a typical strain energy 
curve lying below the one of carbon nanotubes. \footnote{In Ref.~\onlinecite{boustani:99} 
smooth BNTs and a flat BS were compared.} Now, using an \textit{ab initio} method, we also found that armchair BNTs have 
strain energy, but it is higher than for carbon nanotubes.

Different \textit{ideal} zigzag BNTs in Fig.~\ref{fig:strain} have rather low 
strain energies. Here none of the $\sigma$ bonds has to be bent, and the strain 
energy should only come from the multicenter bonds. But those ideal BNTs are 
metastable, and isomers of lower symmetry have higher cohesive energies. Thus 
for the zigzag $\alpha$ isomers no strain energy curve may be plotted as they 
are more or less isoenergetic. It seems that zigzag BNTs can release some or 
all of their strain energy by lowering their symmetry and undergo internal 
deformations (see also  Ref.~\onlinecite{evans_2005_prb}), possibly mediated 
by the jointlike properties of the multicenter bonds.

In summary we see that the strain energy in BNTs is mainly caused by bent 
$\sigma$ bonds lying entirely (armchair) or only partially (chiral BNTs) along 
the circumferential direction. The multicenter bonds are always present, but 
they seem to have no serious effect on this. The apparent \textit{absence} of 
strain energy in zigzag BNTs is caused by the fact that the linear $\sigma$ 
bonds lie along the axial direction, only. But without smoothing bonding
strains, the  zigzag tubes are free to take a \text{multitude} of 
cross-sectional morphologies. This explains the number of different isomers 
that we found for (9,0), (10,0), and (12,0) zigzag BNTs and their bizarre shapes. 
The constriction of zigzag BNT, first reported in 
Ref.~\onlinecite{kunstmann_2005_cpl}, is a clear consequence of the absence 
of strained bonds within zigzag BNTs. Armchair BNTs in turn, which are 
geometrically stabilized by their strain energy, do not seem to have this kind 
of isomerism.

Chiral BNTs may be thought of as a certain combination of structural elements 
from armchair and zigzag tubes defined by a certain chiral angle. Therefore we 
suppose that there will be a separate strain energy curve for every chiral 
angle lying in-between the armchair and the zigzag curves. The strain energies 
themselves will depend on the radii and on the chiral angle of a BNT: 
$E_{\mathrm{strain}}^{\mathrm B}=E_{\mathrm{strain}}^{\mathrm B}(R,\theta)$.
This seems to be a unique property among all nanotubular materials reported so 
far.

But it remains open whether the strain energy of zigzag BNT will be 
completely absent, or just significantly smaller than for armchair BNTs. The 
present results are in favor of the former hypothesis. 
As carbon nanotubes with large diameters (and very small strain energies) are susceptible to a structural collapse, \cite{chopra_1995_nat,elliott_2004_prl} it is possible that without a significant amount of strain energy the zigzag nanotubes could be geometrically \textit{unstable}. Given some thermal fluctuations or strain they might collapse just like big diameter carbon nanotubes. However, such a collapse might also be prevented by a possible energy barrier, which should be proven to be absent in order to support this collapse hypothesis.

Finally we want to point out that the constriction of zigzag BNTs could be 
an important intermediate mechanism during the possible collapse of a zigzag BNT, which might allow for the formation of  B$_{12}$ icosahedra, that are the basic building blocks of all previously known \textit{bulk} boron structures. 
The five-atomic element (see Fig.~\ref{fig:elements}) forms part of an 
imaginary zigzag 6-ring, similar to the six apex atoms of a B$_{12}$ 
icosahedron, as seen along each of its threefold axes. 
\cite{kunstmann_2005_cpl}


\section{Summary and Conclusions}

In this paper we studied a number of different structure models for broad boron 
sheets (BSs). All of them are metallic, and we found that for a 16 atom 
supercell, the model with a simple up-and-down puckering will be the most 
stable one. Large quasiplanar boron clusters with a similar structure
(B$_{22}$, \cite{boust2:97} B$_{48}$, \cite{boustani:98} and
B$_{96}$ \cite{boustani:00}) were already reported before. They may now be 
understood as a first indication for the onset of periodicity in finite 
layered boron systems, and thus they are an independent confirmation of 
the current structure model.

A flat BS has a rather high stiffness, and it seems to be held together 
primarily by multicenter bonds (see Appendix \ref{app:flat}). Although the 
sheet is less stable than previously known bulk phases of boron, as shown 
here and elsewhere, \cite{evans_2005_prb,cabria_2006_nt} the model sheet 
could be the ideal theoretical tool for studying complex multicenter bonds.

After describing the lattice structure of the stable BS, we have analyzed its 
band structure, the corresponding charge densities, and the electron 
localization function. This would lead to the following preliminary picture 
of the chemical bonding: on the one hand the sheet is held together by
homogeneous multicenter bonds, on the other hand there are linear $sp$
hybridized $\sigma$ bonds exclusively lying along the armchair direction of the 
sheet. The existence of $sp$ hybridization in quasiplanar BS is somewhat 
surprising given the fact that earlier studies would always claim $sp^2$ 
hybridization. The rather anisotropic bond properties of the sheets lead 
to different elastic moduli $C_x$ and $C_y$ for stretching the BS in the $x$ 
and in the $y$ direction. Furthermore puckering of the BS, which breaks the 
triangular symmetry, may be understood as a key mechanism to stabilize the 
$sp$ $\sigma$ bonds. Our results indicate that the sheet analyzed in this 
study is the boron analog of a single graphene sheet, a possible precursor 
of boron nanotubes (BNTs), and we wonder whether broad BSs might exist in 
nature. 

Constructing BNTs from the BSs by a cut and paste procedure will generate \textit{ideal}
BNTs (see Appendix \ref{app:idealBNT}). Because the underlying two-dimensional 
lattice structure is rectangular rather than triangular or hexagonal,
it follows that the chiral angle $\theta$ ranges from $0^{\circ}$ to 
$90^{\circ}$ ($\theta = 0^{\circ}$: zigzag, $\theta = 90^{\circ}$: armchair), 
and that chiral BNTs do not have an axial translational symmetry. We therefore 
predict the existence of helical currents in ideal 
chiral BNTs (Appendix \ref{app:Tvec}). Furthermore we presented a band theory for 
ideal BNTs, employing their helical symmetry, and showed that \textit{all} 
ideal BNTs are metallic, irrespective of their radius and chiral angle 
(Appendix \ref{app:bands}). BNTs could therefore be perfect nanowires, superior 
to carbon nanotubes.

In an independent study of armchair and zigzag BNTs we found that ideal BNTs do 
not form the ground state of BNTs, and we identified structures of lower 
symmetry, which are higher in cohesive energy. The symmetries of \textit{real} 
BNTs still remain to be determined, and the ideal BNTs may be seen as rather 
close approximants to real BNTs.

We also found that all BNTs, except small radius armchair types, had puckered 
surfaces, and $\sigma$ bonds along the armchair direction of the primitive 
lattice. The existence and mutual orientation of these $\sigma$ bonds is 
crucial to understand the \textit{basic} mechanical and energetic properties 
of BNTs because the strain energy of the tube is mainly generated by bending 
those $\sigma$ bonds. The multicenter bonds seem to have no real effect on the 
strain energy. They are likely to have jointlike properties (they are easy to 
turn but hard to tear), which allows for a certain flexibility of these bonds, 
and any bonding strain could immediately be released through internal 
relaxations. \cite{evans_2005_prb}

We showed that armchair BNTs, where the $\sigma$ bonds lie along the 
circumferential direction, will have rather high strain energies, whereas 
zigzag BNTs, where the $\sigma$ bonds will lie along their
axial directions, will have nearly vanishing strain energies. Thus BNTs have 
a strain energy that depends on the nanotube's radius $R$ as well as on the 
chiral angle $\theta$:
$E_{\mathrm{strain}}^{\mathrm B}=E_{\mathrm{strain}}^{\mathrm B}(R,\theta)$.
We suppose that there will be an individual strain energy curve for every 
chiral angle lying between the armchair and the zigzag curves. This is a 
unique property among all nanotubular materials reported so far.

This intriguing feature could even allow for some structure control in 
nanotechno\-logy. For carbon nanotubes, the strain energies do not depend on 
their chiralities 
($E_{\mathrm{strain}}^{\mathrm C}=E_{\mathrm{strain}}^{\mathrm C}(R)$), and 
thus the experimentalists may control the radius, but not the chirality of 
carbon nanotubes, although the latter will determine the electronic properties 
of such materials. With the experimental techniques at hand today one might be 
able to walk along the energy axis by tuning the reaction conditions, and 
along the radius axis by synthesizing nanotubes within porous templates 
with well defined pores sizes. This way it could be possible to synthesize 
BNTs of a specific type only. The connection to carbon nanotubes may occur
via intramolecular junctions, \cite{kunstmann_2004_jcp} allowing for a 
controlled layout of carbon nanotubes as well.

The rather low strain energies in \textit{zigzag} BNTs lead to a whole bunch of 
possible structural isomers, as a nanotube without any significant amount of 
strain energy will not be able to maintain a circular cross section. This can 
lead to a certain constriction of zigzag BNTs, \cite{kunstmann_2005_cpl} and 
we even hypothesize that zigzag BNTs could be too unstable to really exist out in nature, provided that there will be no significant energy barrier left to prevent a collapse. 

\textit{Armchair} BNTs on the other hand are geometrically stabilized by their 
strain energies, but for armchair BNTs of rather small radii, the BNTs are 
unable to maintain a puckered structure necessary to align the 
circumferential $\sigma$ bonds. In agreement with earlier studies 
\cite{boustani:97,boustani:98,evans_2005_prb} we expect them to flaten 
out and build up a smooth surface.
Furthermore, we hypothesize an enhanced reactivity of small radius 
armchair BNTs in comparison to zigzag BNTs, which could be useful for 
embedding BNTs into polymers. \cite{evans_2005_prb}

\begin{acknowledgments}
For helpful hints and discussions we thank O. K. Andersen (Stuttgart), 
O. Gunnarsson (Stuttgart), and I. Boustani (Wuppertal). J. Kunstmann 
gratefully acknowledges support from the International Max Planck Research 
School for Advanced Materials (IMPRS-AM). We thank A. Kokalj (Ljubljana) for 
writing the program XCrysDen, \cite{kokalj_2003_cms} that was used to 
generate the atomic structure illustrations.
\end{acknowledgments}


\appendix


\section{\label{app:idealBNT}The Mathematical Description of Ideal Boron 
Nanotubes}

\subsection{\label{app:wvector}Wrapping vector}

\begin{figure}[t]
\centering
\includegraphics[width=\mycwidth]{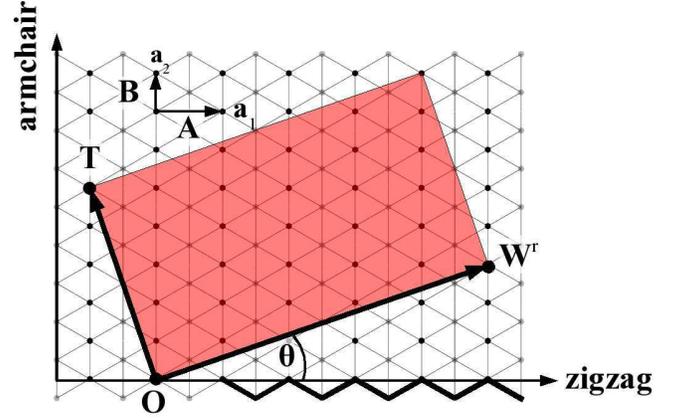}
\caption{\label{fig:cut&roll}(Color online) The geometrical construction of 
an \textit{ideal} boron nanotube from a boron sheet: the red (gray) area is 
cut and rolled up such that $\bm{W}^{\mathrm{r}}$ will become the circumference 
of the nanotube. $\bm{O}$ is the origin, $\bm{W}^{\mathrm{r}}$ is the
wrapping vector, $\bm{T}$ is the translational vector, $\theta$ is the chiral 
angle measured with respect to the zigzag direction, $\bm{a}_{1,2}$ are the 
primitive vectors of the underlying \textit{rectangular} lattice, and $A$ and 
$B$ are the lattice constants (see text). The puckering of the boron sheet is 
indicated by black and gray atoms in the background. The zigzag and the
armchair directions are perpendicular to each other. This figure corresponds 
to $\bm{W}^{\mathrm{r}}=$ (5,3) and $A/B=\sqrt{3}$, which implies 
$\bm{T}= (-1,5)$.}
\end{figure}

The geometrical construction of BNTs from BSs is similar to the 
construction of carbon nanotubes from a graphene sheet: \cite{saito_1989_cnt} 
the basic tubular structure is characterized by a wrapping vector $\bm{W}$ 
that defines a rectangular area on the BS, which is rolled up to a cylinder 
such that $\bm{W}$ becomes the circumference of the nanotube and its radius 
will be $R=|\bm{W}| / 2\pi$ (see Fig.~\ref{fig:cut&roll}). We will call 
any BNT, whose structure may be described by such a construction, an 
\textit{ideal boron nanotube}.

\begin{figure}[t]
\centering
\includegraphics[width=\mycwidth]{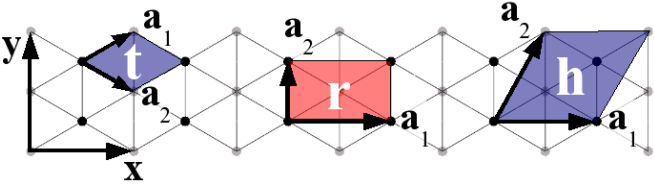}
\caption{\label{fig:cells}(Color online) The triangular (t), the rectangular 
(r), and the honeycomb-derived (h) primitive cells that are used to 
characterize boron nanotubes. They contain one, two, and three atoms, 
respectively. Only the rectangular cell may properly describe the puckering 
of the boron sheet (indicated by black and gray atoms in the background).}
\end{figure}

Due to the fact that a proper structure model for BS was missing for a long 
time, there remains some confusion in the literature about a proper reference 
lattice structure. In the work of Cabria \textit{et al.} \cite{cabria_2006_nt} 
and in earlier works by us \cite{kunstmann_2005_cpl,boustani:99} (and in 
full analogy to the construction of carbon nanotubes) the BNTs are related 
to a honeycomb lattice and defined the wrapping vector $\bm{W}^{\mathrm{h}}$ 
as
\begin{equation}
\bm{W}^{\mathrm{h}} = (n,m)= n \bm{a}_1^{\mathrm{h}} + m \bm{a}_2^{\mathrm{h}},
\end{equation}
$\bm{a}_{1,2}^{\mathrm{h}}$ are the primitive vectors of a honeycomb lattice 
and $n,m$ are integers. Here each unit cell has one additional atom at the 
center of the honeycombs, thus consisting of three rather than two atoms 
(see Fig.~\ref{fig:cells}).
Gindulyte \textit{et al.}, \cite{gindulyte_1998_ic1} Evans 
\textit{et al.}, \cite{evans_2005_prb} and some earlier work of ours 
\cite{boustani:99} relate their BNTs to the simple 
triangular lattice, having only one atom per unit cell:
\begin{equation}
\bm{W}^{\mathrm{t}} = (i,j)= i \bm{a}_1^{\mathrm{t}} + j 
\bm{a}_2^{\mathrm{t}},
\end{equation}
$\bm{a}_{1,2}^{\mathrm{t}}$ are the primitive vectors of a triangular 
lattice, and $i,j$ are integers. $\bm{W}^{\mathrm{h}}$ and 
$\bm{W}^{\mathrm{t}}$ can be transformed into each other
by using \footnote{These relations are obtained from: 
$\bm{a}_1^{\mathrm{h}} = \bm{a}_1^{\mathrm{t}}
+ \bm{a}_2^{\mathrm{t}}$ and $\bm{a}_2^{\mathrm{h}} = 
2\bm{a}_1^{\mathrm{t}} - \bm{a}_2^{\mathrm{t}}$.}
\begin{eqnarray}
 (n,m) &\mapsto& (i,j)=(n+2m,\ n-m),\\
 (i,j) &\mapsto& (n,m)=\frac{1}{3} (i+2j,\ i-j).
\end{eqnarray}
From Fig.~\ref{fig:cells} we see that both definitions are based 
on primitive vectors, which have different orientations. 
\footnote{For the plot in Fig.~\ref{fig:cells} we used $A=\sqrt{3}B$.
This induces triangular symmetry into a rectangular lattice, and the 
primitive vectors are 
$\bm{a}_{1}^{\mathrm{h}}=\bm{a}_1^{\mathrm{r}}=\sqrt{3}B(1,0)$,
$\bm{a}_{2}^{\mathrm{h}}=\sqrt{3}B(\frac{1}{2},\frac{1}{2} \sqrt{3})$, and
$\bm{a}_1^{\mathrm{t}}= B(\frac{\sqrt{3}}{2},\frac{1}{2})$, 
$\bm{a}_2^{\mathrm{t}}=
B(\frac{\sqrt{3}}{2},-\frac{1}{2})$.}
This leads to the rather unsatisfactory situation that armchair and zigzag 
directions are swapped in both descriptions (see Table \ref{tab:BNT} for 
example). Cabria \textit{et al.} found that all $(n,0)$ zigzag and all 
$(2n,2n)$ armchair BNTs have puckered surfaces, while the $(2n+1,2n+1)$ 
armchair tubes shall be smooth due to the fact that an odd number of boron 
rows along the tube surfaces does not allow for the formation of the simple 
up and down puckering. \cite{cabria_2006_nt} We think that these results 
are not an intrinsic property of BNTs, but rather a consequence of an unsuitable 
reference lattice system that is unable to properly describe the puckering of 
the boron sheet, see Fig.~\ref{fig:cells}. Furthermore, the puckering will break 
the hexagonal symmetry underlying the honeycomb and the triangular lattice.

Therefore we convinced ourselves that these descriptions are not really 
appropriate to classify BNTs. On the basis of the current BS model we would like to put 
forward a \textit{different} way of describing BNTs, based on a rectangular 
lattice underlying the two-dimensional structure of the BS.

We define the wrapping vector $\bm{W}^{\mathrm{r}}$ as
\begin{equation}
\bm{W}^{\mathrm{r}} = (k,l)= k \bm{a}_1^{\mathrm{r}} + 
l \bm{a}_2^{\mathrm{r}},
\label{eqn:W}
\end{equation}
$k,l$ are integers, and $\bm{a}_1^{\mathrm{r}}=A(1,0)$ and 
$\bm{a}_2^{\mathrm{r}}=B(0,1)$ are the primitive vectors of the rectangular 
lattice (see Figs.~\ref{fig:cells} and \ref{fig:cell}); $A$
and $B$ are the lattice constants from Table \ref{tab:lattice}. In analogy 
to the Dresselhaus construction for carbon nanotubes, \cite{saito_1989_cnt} 
we define the chiral angle $\theta$ as the angle between the vectors 
$\bm{W}^{\mathrm{r}}$ and $\bm{a}_1^{\mathrm{r}}$, i.e., $\theta$
is measured with respect to the zigzag direction coinciding with 
$\bm{a}_1^{\mathrm{r}}$
(see Fig.~\ref{fig:cut&roll}).

The categorization of BNTs will be different from other classification schemes 
because the reduced symmetry of a BS increases the number of possible types of 
nano\-tubes, as the range for the chiral angle will be $0^{\circ} \le \theta 
\le 90^{\circ}$, and for the chiral indices $(k,l)$ we find that $k,l \ge 0$. 
Zigzag BNTs will now correspond to $\theta = 0^{\circ}$ and $(k,l)=(k,0)$, and 
armchair BNTs will correspond to $\theta = 90^{\circ}$ and $(k,l)=(0,l)$.

$\bm{W}^{\mathrm{h}}$ and $\bm{W}^{\mathrm{t}}$ cannot directly be converted to
$\bm{W}^{\mathrm{r}}$, as they are defined for lattices with different symmetries. 
For the achiral types, on may use the following
analogy (for examples see Table \ref{tab:BNT})
\begin{eqnarray}
\hbox{zigzag:} \quad (k,0)^r &\leftrightarrow& (k,0)^h \\
&\leftrightarrow& (k,k)^t, \nonumber\\
\hbox{armchair:} \quad (0,l)^r &\leftrightarrow& (l/3,l/3)^h \nonumber\\
&\leftrightarrow& (l,0)^t. \nonumber
\end{eqnarray}


\subsection{\label{app:Tvec}Translational vector}

The \textit{tubular} unit cell of an ideal BNT, being the red (gray) area in 
Fig.~\ref{fig:cut&roll}, may be defined properly by a wrapping vector 
$\bm{W}^{\mathrm{r}}$ (Eq.~(\ref{eqn:W})) and the so-called
translational vector $\bm T$, which is perpendicular to 
$\bm{W}^{\mathrm{r}}$:
\begin{eqnarray}
\bm{T} = (t_1,t_2) &=& t_1 \bm{a}_1^{\mathrm{r}} + 
t_2 \bm{a}_2^{\mathrm{r}} \label{eqn:T},\\
t_1 &=&
   \left\{ \begin{array}{c@{\quad:\quad}l}
      -\hbox{numerator}(f) & k \ne 0 \\
      1 & k=0\\
   \end{array} \right. \nonumber \\
t_2 &=&
   \left\{ \begin{array}{c@{\quad:\quad}l}
      \hbox{denominator}(f)\nonumber & k \ne 0\\
      0 & k=0\\
   \end{array} \right. \nonumber \\
f &=& \hbox{reduce}\left( \frac{lB^2}{kA^2}\right) \nonumber.
\end{eqnarray}
$t_1,t_2$ are integers and reduce ($r$) should indicate that the
fraction $r$ must be reduced before determining its numerator and
denominator.

\begin{figure}[t]
\centering
\includegraphics[width=0.7\mycwidth]{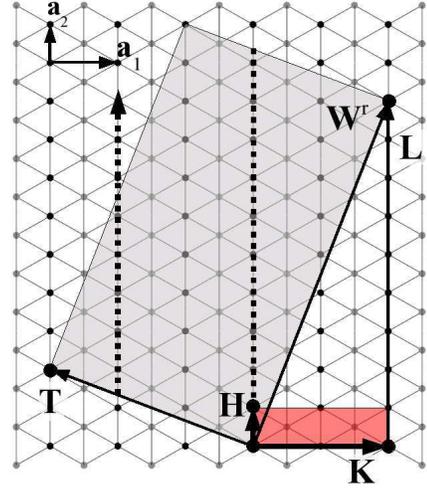}
\caption{\label{fig:helical}(Color online) Two different ways of "building up" 
a nanotube: the \textit{tubular} unit cell in light gray 
(see also Fig.~\ref{fig:cut&roll}) is repeated along the nanotube's axis,
which lies parallel to $\bm{T}$. The \textit{helical} unit cell in red 
(dark gray) is translated along spirals (represented by the dotted lines) on 
the surface of the nanotube; it is defined by the helical vector $\bm{H}$ 
and vector $\bm{K}$. It holds $\bm{W}^{\mathrm{r}}=\bm{K} + \bm{L}$.
Here $\bm{H}=(0,1)$, $\bm{K}=(2,0)$ and $\bm{L}=(0,9)$, and therefore 
$\bm{W}^{\mathrm{r}}=(2,9)$. The length of $\bm{T}=(-3,2)$ was artificially 
reduced by choosing $A/B=\sqrt{3}$.}
\end{figure}

Let us consider the length of the translational vector $\bm T$.
For the achiral BNTs $|\bm T|$ is particularly small: for all $(k,0)$ zigzag 
types we have $\bm{T}=(0,1)$, and for $(0,l)$ armchair BNTs $\bm{T}=(1,0)$.
For the chiral types $\bm T$ depends on the ratio $B^2/A^2$ (see the last line 
of Eq.~(\ref{eqn:T})). Using $A=2.819$ and $B=1.602$ we obtain reduce 
$(B^2/A^2)=2 566 404/7 946 761$. Therefore the coefficients
$t_1$ and $t_2$ are really huge numbers, which means that $|\bm T|$ becomes 
macroscopically large.
For $A$ and $B$ chosen as above, the estimated length of $\bm T$ for all 
chiral BNTs will be in the mm range. Imposing some additional symmetry 
constraints by relating the lattice constants will immediately remedy this 
problem. For example by choosing $A=\sqrt{3}B$, fraction$(B^2/A^2)=1/3$,
i.e., $|\bm T|$ will be reduced to just a few lattice constants (this case 
was used to generate Figs.~\ref{fig:cut&roll} and \ref{fig:helical}).
So for the chiral BNTs the specific ratio $B^2/A^2$ determines the length 
of the translational vector.

Boron compounds usually have a whole set of different B--B bond lengths, which 
means that boron does not necessarily favor highly symmetric arrangements. The 
bond lengths are more flexible than for typical covalent elements like carbon, 
and the lattice constants $A$ and $B$ of the BS cannot really be seen as fixed 
parameters; they will have slightly different values in BNTs. Furthermore, the 
broken planar triangular symmetry of the BS is rather typical for boron, and 
we should expect that for ideal chiral BNTs, even with different values of 
$A$ and $B$, the translational vector might still be large.

To summarize: any departure from triangular symmetry in the BS will create 
chiral BNTs, which have macroscopically large translational vectors, and 
achiral types, where $|\bm T|$ is of the order of the lattice constants. Thus 
achiral BNTs (armchair and zigzag) have a one-dimensional translational 
symmetry along the tube's axis, which is not present in chiral BNTs. For the 
latter it might be better to think in terms of helical (chiral) symmetries only. 
Therefore we predict the existence of \textit{helical currents} in ideal chiral 
BNTs. Such currents could lead to very interesting physical effects such as 
strong magnetic fields \cite{bagci_2002_prb} and self-inductance effects 
leading to an inductive reactance \cite{miyamoto_1999_prb} of chiral BNTs.

\subsection{\label{app:bands}Band structure}

Within the limit of large nanotube radii, where curvature effects are small, 
one may derive the one-dimensional band structure of an ideal BNT 
$E_{\mu}(k^{\prime})$ by a zone-folding technique,
\cite{saito_1989_cnt} starting from the two-dimensional band structure of a BS
$E^{\mathrm{BS}}(k_x,k_y)$. Given the absence of translational symmetry in ideal 
chiral BNTs, we have to base our zone-folding theory on the helical symmetry of 
BNTs. \cite{white_1993_prb, lin-chung_1994_jpcm}

Figure \ref{fig:helical} illustrates that besides constructing a BNT by repeating 
a \textit{tubular} unit cell one can also build a nano\-tube by repeating a 
\textit{helical} unit cell along a spiral winding around the surface of the 
tube. The direction of this spiral is given by the helical vector $\bm{H}$ 
\cite{white_1993_prb, lin-chung_1994_jpcm} 
(in Ref.~\onlinecite{saito_1989_cnt} it is called the symmetry vector 
$\bm R$), which, when uncoiled into a plane, defines the direction
of a translational symmetry (see Eq.~(\ref{eqn:bloch1}) and thereafter). The 
helical unit cell is specified by $\bm H$ and the vector $\bm{K} \perp \bm{H}$. 
Furthermore, we define a vector $\bm{L} \parallel \bm{H}$, such that 
$\bm{W}^{\mathrm{r}}= (k,l) = \bm{K} + \bm{L}$
(see Fig.~\ref{fig:helical}).

The helical wave functions are restricted by the following criteria:
\begin{eqnarray}
\Psi_{\mu k^{\prime}}(\bm{r} + \bm{H}) &=& \Psi_{\mu k^{\prime}}(\bm{r}) 
\exp(ik^{\prime}|\bm{H}|) \label{eqn:bloch1},\\
\Psi_{\mu k^{\prime}}(\bm{r}+\bm{W}^{\mathrm{r}}) &=& 
\Psi_{\mu k^{\prime}}(\bm{r}) \label{eqn:RB1}.
\end{eqnarray}
Equation (\ref{eqn:bloch1}) defines a one-dimensional Bloch state
with $-\pi /|\bm{H}| < k^{\prime} < \pi /|\bm{H}|$ and imposes the 
condition that $k^{\prime}$ has to be parallel to the reciprocal vector 
of $\bm{H}$. Equation (\ref{eqn:RB1}) is the tubular boundary condition. In order 
to construct the helical wave functions $\Psi_{\mu k^{\prime}}$ we have to 
use the wave functions of the BS $\Psi_{\bm{k}}^{\mathrm{BS}}(\bm{r})$
which have the Bloch property:
\begin{equation}
\Psi_{\bm{k}}^{\mathrm{BS}}(\bm{r}+\bm{R}) = \exp(i\bm{k} \cdot \bm{R}) 
\Psi_{\bm{k}}^{\mathrm{BS}}(\bm{r}),
\label{eqn:bloch2}
\end{equation}
where $\bm{R}$ is a vector of the Bravais lattice formed by 
$\bm{a}^{\mathrm{r}}_1$ and $\bm{a}^{\mathrm{r}}_2$. Since the vectors 
$\bm{H}$ and $\bm{W}^{\mathrm{r}}$ are also elements of such a
Bravais lattice, Eq.~(\ref{eqn:bloch1}) will automatically be satisfied, 
and Eq.~(\ref{eqn:RB1}) together with Eq.~(\ref{eqn:bloch2}) will yield
\begin{equation}
1=\exp \left[ i(\bm{k} \cdot \bm{W}^{\mathrm{r}}) \right].
\label{eqn:RB2}
\end{equation}

In order to proceed, we now have to define the direction of $\bm{H}$, which 
may be any Bravais lattice vector.
\footnote{In the most general case the zone-folded band structure of ideal 
BNTs is given by $E_{\mu}(k^{\prime}) = 
E^{\mathrm{BS}}(k^{\prime}\bm{G_H}/|\bm{G_H}| + \mu \bm{G_K})$,
with $\bm{G_H}$ and $\bm{G_K}$ being the reciprocal lattice vectors 
of $\bm{H}$ and $\bm{K}$,
respectively, $-\pi /|\bm{H}| < k^{\prime} < \pi /|\bm{H}|$, $\mu = 0, \dots, N-1$, 
and $N=|\bm{H} \times \bm{K}|/ |\bm{a}_1^{\mathrm{r}} \times 
\bm{a}_2^{\mathrm{r}}|$. \cite{saito_1989_cnt}} By choosing $\bm{H}=\bm{T}$ 
we recover the case of a tubular unit cell, as described above and in
Ref.~\onlinecite{saito_1989_cnt}. But in order to make the calculation 
as simple as possible we assign
$\bm{H}=\bm{a}^{\mathrm{r}}_2=(0,1)$. Then it follows that
$\bm{K}=(k,0)$ and $\bm{L}=(0,l)$ (see Fig.~\ref{fig:helical}). As $\bm{H} 
\parallel y$ we have to choose $k^{\prime}=k_y$. After inserting 
Eq.~(\ref{eqn:RB2}) into $E^{\mathrm{BS}}(k_x,k_y)$ we finally obtain the 
zone-folded band structure of ideal $(k,l)$ ($k \ne 0$) BNTs as
\begin{eqnarray}
E_{\mu}^{(k,l)}(k^{\prime}) = E^{\mathrm{BS}}
\left( \frac{2\pi}{kA} \mu - \frac{lB}{kA} k^{\prime}, k^{\prime} \right) 
\label{eqn:bands1}, \\
-\frac{\pi}{B} < k^{\prime} < \frac{\pi}{B}, \nonumber\\
\mu = 0, \cdots, k-1 \nonumber
\end{eqnarray}

Equation (\ref{eqn:bands1}) will break down for $(0,l)$ armchair BNTs, due 
to a chiral index $k=0$. But as mentioned before, we are free to choose 
the direction of $\bm{H}$, and in such a case we use
$\bm{H}=\bm{a}^{\mathrm{r}}_1=(1,0)$ and have $k^{\prime}=k_x$. We 
thus obtain
\begin{eqnarray}
E_{\mu}^{(0,l)}(k^{\prime}) = E^{\mathrm{BS}}\left( k^{\prime}, 
\frac{\mu}{l}\frac{2\pi}{B} \right)
\label{eqn:bands2},\\
-\frac{\pi}{A} < k^{\prime} < \frac{\pi}{A} \nonumber,\\
\mu = 0, \cdots, l-1 \nonumber
\end{eqnarray}

Unfortunately we do not have an analytical band structure of the BS 
$E^{\mathrm{BS}}(k_x,k_y)$, yet. But to decide whether a certain ideal 
BNT is metallic or not we can simply zone-fold the BS's Fermi surface 
given in Fig.~\ref{fig:FS}. We did so and found that \textit{all} ideal 
BNTs are indeed metallic, irrespective of their radius and chiral angle. 
The only ideal BNTs that are not metallic are the (0,1) and the (0,2) types. 
But these structures are highly unrealistic and we can safely rule them out, 
as they are not even covered by the Aufbau principle. 
\cite{boustani_1997_prb}



\begin{figure}[t]
\setlength{\myw}{5.1cm} 
\centering
\subfigure(a){\includegraphics[width=0.4\mycwidth]{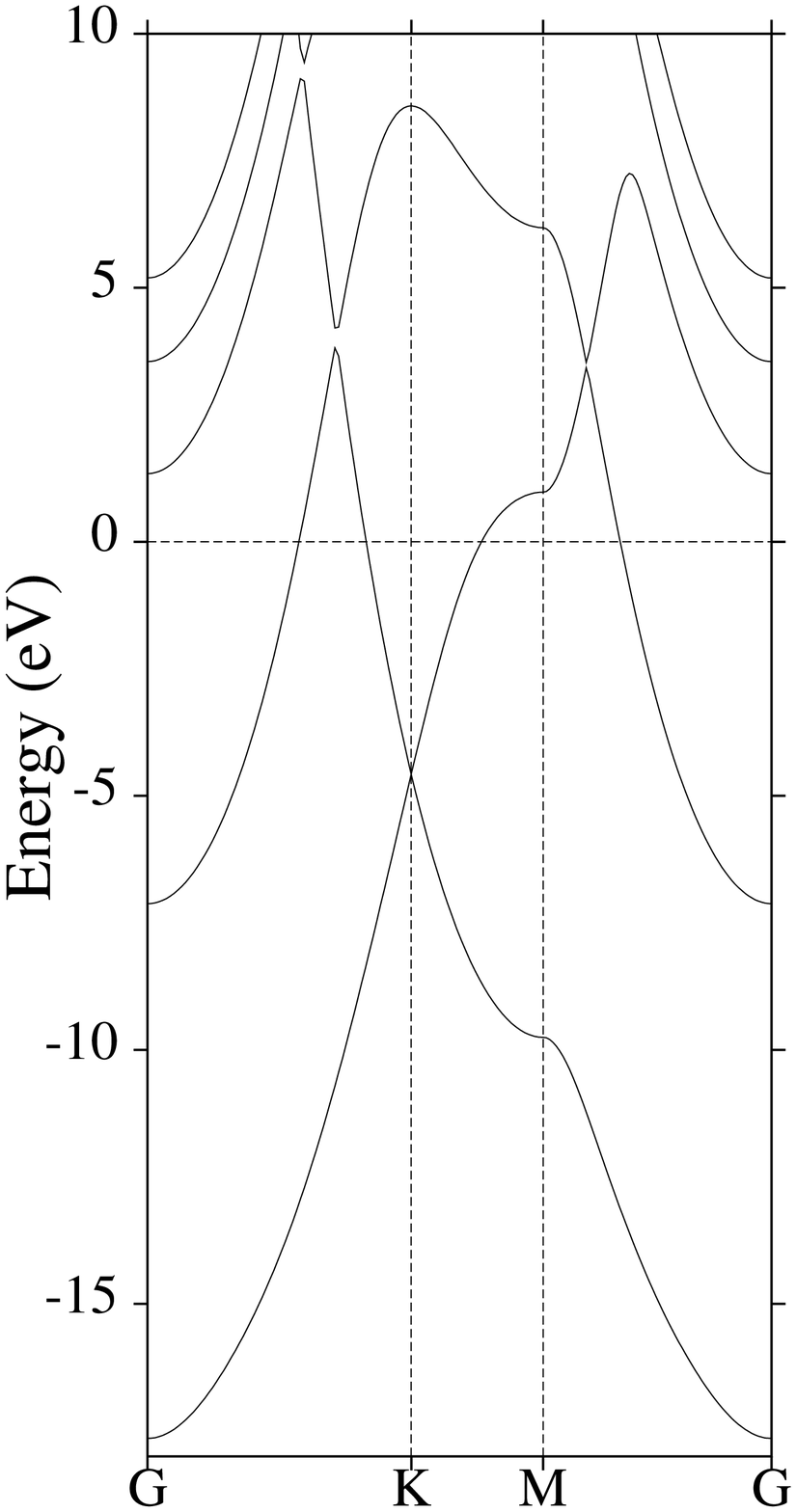}}
\subfigure(b){\includegraphics[width=0.45\mycwidth]{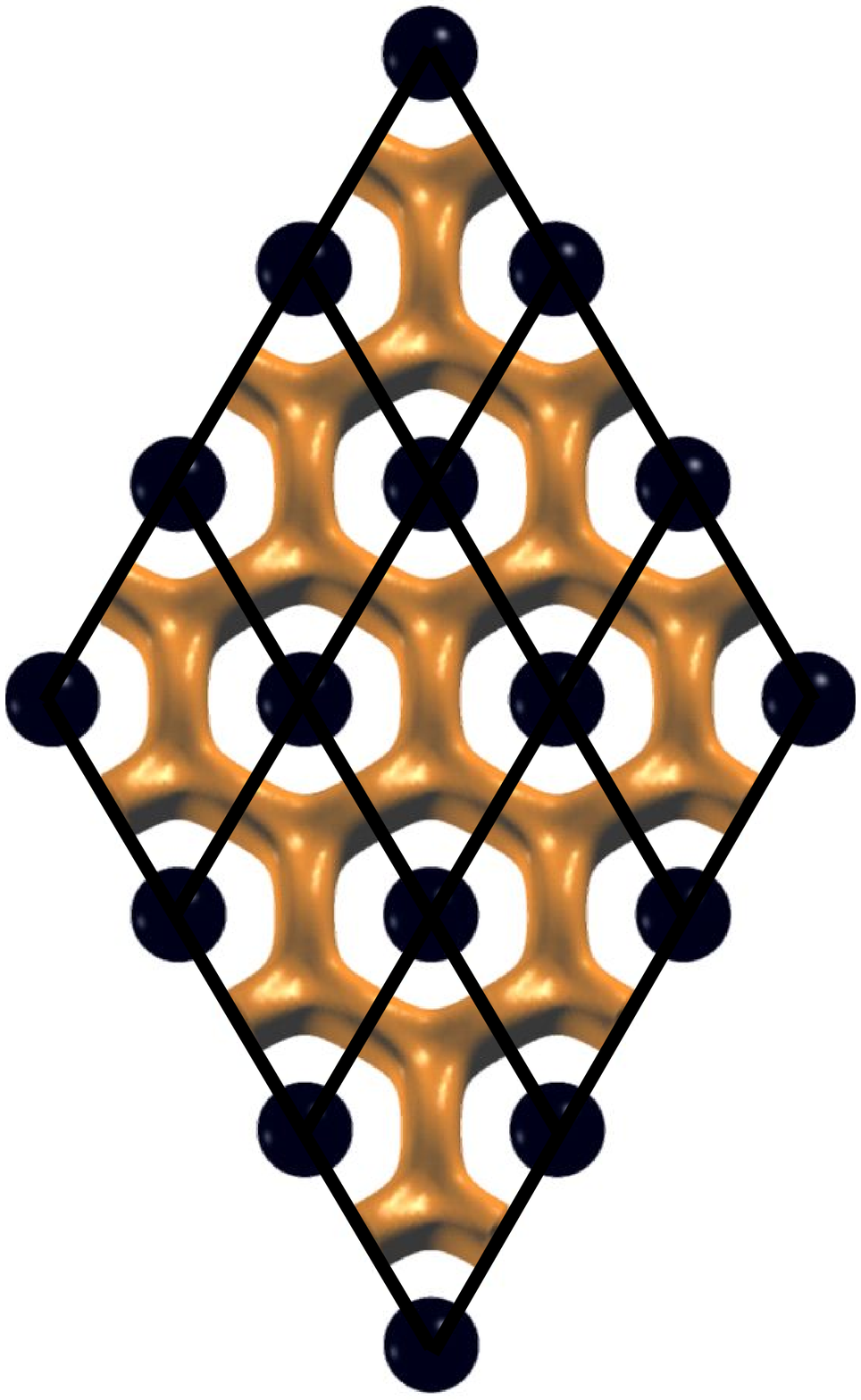}}
\caption{\label{fig:flat}(Color online) Properties of a flat boron sheet: 
(a) The two-dimensional band structure. (b) Black lines indicate the 
triangular unit cells, black spheres are boron atoms, and the 
orange (gray) contours show the electron localization function (ELF) 
at contours of 0.7. We observe a simple network of two- and three-center 
bonds.}
\end{figure}

\begin{figure}[t]
\setlength{\myw}{5.1cm} 
\centering
\includegraphics[width=0.4\linewidth]{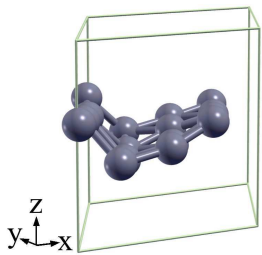}
\caption{\label{fig:kink}(Color online)
A kinked boron sheet based on a supercell (thin lines) that contains 
16 atoms (see text). Apart from the kink its surface is slightly puckered.
}
\end{figure}


\section{\label{app:flat}A Flat Boron Sheet}

The lattice structure, the cohesive energy, and the elastic moduli of the flat 
boron sheet -- model (a) -- can be found in Table \ref{tab:lattice}. The elastic 
modulus of $C_{11} \approx 750$ GPa is comparable to graphite. The electronic 
charge density is nearly uniform in the interstitial region, and the band 
structure (see Fig.~\ref{fig:flat}(a)) is similar to the band structure of 
a free electron gas. These results seem to indicate some metallic bonding, as 
pointed out by Evans \textit{et al.}, \cite{evans_2005_prb} but such a 
picture cannot really account for the planarity and the high elastic modulus 
of the flat BS. A different qualitative picture of the chemical bonding is 
obtained after looking at the electron localization function 
\cite{becke_1990_jcp} (ELF) in Fig.~\ref{fig:flat}(b).
Here we observe a simple network of two- and three-center bonds being less
localized (ELF $\approx 0.7$) than typical sigma bonds (ELF $\approx 0.9$), 
which are absent here. Thus the flat BS seems to be held together predominantly 
by multicenter bonds similar to the ones found in pure boron compounds. The 
chemical understanding of these multicenter bonds is still very limited. We 
think that, despite of its apparent metastability, model (a) could be an ideal 
theoretical tool to extend our present understanding of the nature of multicenter 
bonding in boron.

\section{\label{app:kink} A Kinked Boron Sheet}

In Sec.~\ref{sec:model} we described the optimization of randomly puckered 
BSs. It is surprising that despite the high complexity of the boron energy 
landscape for small boron clusters, which are known to have many local minima, 
these runs seem to have only two possible "attractors". One is model (b) -- a 
simply puckered BS -- the other is the kinked BS displayed in Fig.~\ref{fig:kink}. 
The kinked BS has a metallic density of states and a cohesive energy of 6.86 eV/atom, 
which is somewhat intermediate between model (a) and model (b). We think that this 
structure is likely to be an artifact of the finite size of the supercell that we used 
for the simulation runs, but being an "attractor" of the optimization runs it is 
still interesting enough to be mentioned here.

\bibliography{ihsan+alex,boron,clusters,bulk,own,numerics,general_cnt,misc,elec}

\end{document}